# RIS-Driven Resource Allocation Strategies for Diverse Network Environments: A Comprehensive Review

Manzoor Ahmed, Fang Xu, Yuanlin Lyu, Aized Amin Soofi, Yongxiao Li, Feroz Khan, Wali Ullah Khan, Muhammad Sheraz, Teong Chee Chuah, and Min Deng

*Abstract*—This comprehensive survey examines how Reconfigurable Intelligent Surfaces (RIS) revolutionize resource allocation in various network frameworks. It begins by establishing a theoretical foundation with an overview of RIS technologies, including passive RIS, active RIS, and Simultaneously Transmitting and Reflecting RIS (STAR-RIS). The core of the survey focuses on RIS's role in optimizing resource allocation within Single-Input Multiple-Output (SIMO), Multiple-Input Single-Output (MISO), and Multiple-Input Multiple-Output (MIMO) systems. It further explores RIS integration in complex network environments, such as Heterogeneous Wireless Networks (HetNets) and Non-Orthogonal Multiple Access (NOMA) frameworks. Additionally, the survey investigates RIS applications in advanced communication domains like Terahertz (THz) networks, Vehicular Communication (VC), and Unmanned Aerial Vehicle (UAV) communications, highlighting the synergy between RIS and Artificial Intelligence (AI) for enhanced network efficiency. Summary tables provide comparative insights into various schemes. The survey concludes with lessons learned, future research directions, and challenges, emphasizing critical open issues.

Beyond 5G and 6GReconfigurable Intelligent SurfacesIntelligent Reflecting SurfacesSmart radio environmentAIDRL

## I. INTRODUCTION

In the rapidly evolving landscape of wireless communications, optimizing resource allocation is paramount for achieving optimal system performance. This multifaceted endeavor

Fang Xu, Manzoor Ahmed, and Min Deng are with the School of Computer and Information Science and also with the Institute for AI Industrial Technology Research, Hubei Engineering University, Xiaogan, 432000, China, Fang Xu is also with the School of Computer Science and Information Engineering, Hubei University, Wuhan 430062, China (e-mails: xf2012@whu.edu.cn, manzoor.achakzai@gmail.com, dengmin83@hbeu.edu.cn).

Yuanlin Lyu is with the School of Computer Science and Information Engineering, Hubei University, Wuhan, 430062, China, and also with the College of Technology, Hubei Engineering University, Xiaogan, 432000, China (e-mail: lyl2021@stu.hubu.edu.cn).

Aized Amin Soofi is with the Department of Computer Science, National University of Modern Languages Faisalabad, 38000, Pakistan (e-mail:aizedamin@gmail.com ).

Yongxiao Li is with the Electronics Engineering Department, Beijing University of Posts and Telecommunications, Beijing, 100876, China (e-mail: Yongxiao.li@bupt.edu.cn).

Feroz Khan is with the Balochistan University of Information Technology Engineering & Management Sciences Quetta, Pakistan (e-mail: ferozkhan687@gmail.com).

Wali Ullah Khan is with the Interdisciplinary Centre for Security, Reliability, and Trust (SnT), University of Luxembourg, 1855 Luxembourg City, Luxembourg (e-mails: waliullah.khan@uni.lu).

Muhammad Sheraz and Teong Chee Chuah are with with the Centre for Wireless Technology, Faculty of Engineering, Multimedia University, Cyberjaya Selangor 63100, Malaysia (emails: engr.msheraz@gmail.com, tc-chuah@mmu.edu.my).

TABLE I: Summary of Important Acronyms

| Acronym | Definition |
|---|---|
| 5G | Fifth Generation |
| 6G | Sixth Generation |
| AI | Artificial Intelligence |
| AoI | Age of Information |
| AO | Alternating Optimization |
| BCD | Block Coordinate Descent |
| BS | Base Station |
| CSI | Channel State Information |
| DDPG | Deep Deterministic Policy Gradient |
| DDQN | Double Deep Q-Network |
| DRL | Deep Reinforcement Learning |
| D2D | Device-to-Device |
| EE | Energy Efficiency |
| EMF | Electromagnetic Fields |
| EMI | Ergodic Mutual Information |
| FD | Full-Duplex |
| FL | Federated Learning |
| FSL | Federated Spectrum Learning |
| HD | Half-Duplex |
| HetNets | Heterogeneous Wireless Networks |
| IoT | Internet of Things |
| IOS | Intelligent Omni-Surface |
| ISAC | Integrated Sensing and Communication |
| LoS | Line of Sight |
| MIMO | Multiple-Input Multiple-Output |
| MISO | Multiple-Input Single-Output |
| MDP | Markov Decision Process |
| MEC | Mobile Edge Computing |
| ML | Machine Learning |
| MSE | Mean Square Error |
| NOMA | Non-Orthogonal Multiple Access |
| NN | Neural Networks |
| OFDMA | Orthogonal Frequency-Division Multiple Access |
| OFDM | Orthogonal Frequency Division Multiplexing |
| OMA | Orthogonal Multiple Access |
| PLS | Physical Layer Security |
| QoS | Quality of Service |
| RL | Reinforcement Learning |
| RIS | Reconfigurable Intelligent Surfaces |
| SCA | Successive Convex Approximation |
| SCMA | Sparse Code Multiple Access |
| SDR | Semi-Definite Relaxation |
| SIMO | Single-Input Multiple-Output |
| SINR | Signal-to-Interference and Noise Ratio |
| SNR | Signal-to-Noise Ratio |
| SSR | System Sum Rate |
| STAR | Simultaneously Transmitting and Reflecting RIS |

TABLE I: Summary of Important Acronyms (Continued)

| IX | |
| --- | --- |
| **Acronym** | **Definition** |
| SWIPT | Simultaneous Wireless Information and Power Transfer |
| THz | Terahertz |
| UAV | Unmanned Aerial Vehicle |
| URLLC | Ultra-Reliable Low-Latency Communication |
| VC | Vehicular Communication |
| VLC | Visible Light Communication |
| V2I | Vehicle-to-Infrastructure |
| V2V | Vehicle-to-Vehicle |
| V2X | Vehicle-to-Everything |
| VR | Virtual Reality |
| WMMSE | Weighted Minimum Mean Square Error |
| WNs | Wireless Networks |
| WSR | Weighted Sum Rate |

involves managing a spectrum of resources including power, bandwidth, time slots, and spatial channels, especially in the realm of multi-antenna systems [1]. Successful resource allocation strategies rely heavily on access to comprehensive system data, encompassing vital metrics such as Channel State Information (CSI), queue states, and Quality of Service (QoS) requirements. Numerous studies have underscored the pivotal role of resource allocation in meeting diverse operational objectives, ranging from maximizing system throughput to enhancing Energy Efficiency (EE) and minimizing power consumption. Consequently, there is a pressing need for adaptive strategies tailored to the specific demands of various network scenarios [2]. Table I presents a comprehensive list of all abbreviations pertinent to the subject before continuing.

This study aims to explore how Reconfigurable Intelligent Surfaces (RIS) are revolutionizing resource allocation across a spectrum of network environments, thereby significantly enhancing communication efficiency and system adaptability. RIS, with its capacity to actively modify the radio environment, fundamentally alters the utilization of conventional resources like spectrum, power, and spatial channels [3]. Comprising numerous small elements capable of independently adjusting the phase and amplitude of incident electromagnetic waves, RIS requires dynamic reconfiguration to optimize signal reception, minimize interference, and enhance overall network performance in real-time. While passive RIS merely reflects and modulates incoming signals without consuming power, active RIS requires energy for signal amplification, necessitating meticulous power management to strike a balance between performance enhancements and energy consumption [4].

Moreover, RIS enables precise spatial management through sophisticated beamforming techniques by adjusting phase shifts to direct energy where most needed, thereby reducing interference and enhancing signal coverage. To fully leverage the benefits of RIS, it must be seamlessly integrated with existing network control systems, requiring synchronization with Base Station (BS) and other network components for strategic reconfiguration. This integration necessitates the sharing of real-time CSI, crucial for optimizing RIS configuration. However, the accurate acquisition of CSI is challenging due to the influence of RIS on signal paths, requiring innovative approaches for rapid and efficient updates [5].

RIS is revolutionizing resource allocation strategies within various network configurations, significantly boosting system efficiency and signal quality. In complex and dense networks such as Single-Input Multiple-Output (SIMO), Multiple-Input Single-Output (MISO), and Multiple-Input Multiple-Output (MIMO) systems, RIS effectively optimizes resource allocation by manipulating the propagation environment to enhance signal reception and minimize interference. In Heterogeneous Wireless Networks (HetNets), RIS dynamically adjusts its properties to match different network standards and protocols, improving spectrum and power management to enhance QoS [6]. For Non-Orthogonal Multiple Access (NOMA) systems, RIS refines power distribution and boosts signal discrimination at receivers, enabling more efficient user multiplexing on the same frequency band [7].

RIS also plays a pivotal role in Terahertz (THz) communication networks, which face challenges due to high propagation losses over long distances. Here, RIS significantly extends signal range and reliability [8], [9]. In Vehicular Communication (VC) and Unmanned Aerial Vehicle (UAV) communications, RIS ensures robust connectivity, which is essential for safety and reliability by mitigating rapid changes in Line of Sight (LoS) conditions [10]. Furthermore, edge networks benefit from RIS by enhancing connectivity, reducing latency, and optimizing bandwidth usage through localized signal adjustments [11]. The integration of Artificial Intelligence (AI) with RIS technologies marks a significant step toward automated and intelligent network management. By leveraging AI for predictive analytics and Machine Learning (ML), network parameters are dynamically adjusted in real-time, ensuring optimal performance. This synergy between AI and RIS is paving the way for more adaptive, efficient, and responsive wireless communications, demonstrating the transformative potential of RIS in modern network scenarios [12].

### A. Related Surveys

Existing literature extensively covers RIS from diverse perspectives. However, this survey stands out by offering a cohesive thematic focus. Our motivation is rooted in the exploration of how RIS-driven resource management can significantly enhance communication efficiency and system adaptability across varied network architectures and communication scenarios. The thematic clarity of our paper becomes evident when considering previous research contributions presented.

For instance, [13] examined and compared the effectiveness, hardware requirements, and practical uses of RIS against backscatter and relay technologies. [14] introduced RIS for improving signal reflection efficiency, detailing its functions and summarizing recent research in the field. [15] provided an analysis from a communication theory standpoint on critical technologies for Wireless Networks (WNs) supported by RIS, while highlighting major research areas in this domain. [16] analyzed recent applications and design features of RISs in WNs and explored their innovative implementations. [17] described a future concept for intelligent urban areas in emerging communication networks, focusing on unique application scenarios and use cases. This highlighted the prospective benefits and exciting research possibilities of deploying RIS. [18] outlined different hardware approaches for achieving signal reflection efficiency, analyzed channel characteristics, and investigated the difficulties and potential developments in WNs enhanced by RIS. [19] presented guidance on wireless communications improved through RIS, addressing crucial technical matters from a communications viewpoint. [20] summarized wireless communications assisted by RIS, covering its principles, uses, main performance metrics, as well as upcoming obstacles and strategies for implementation. [21] explored potential applications and designs of RIS devices in different Internet of Things (IoT) sectors and emergency networks, evaluating enhancements in signal quality and identifying future research hurdles. [22] outlined solutions for CSI acquisition, passive information transfer, and phase shift optimization in RIS-enhanced WNs. [23] demonstrated a prototype along with its experimental outcomes, and offered a tutorial on sensing and localization aided by RIS, tackling crucial technical challenges. [24] presented an overview of the fundamentals of RIS and the latest research from a signal processing perspective, encompassing aspects of communication, localization, and sensing. [25] performed a thorough examination of wireless communications assisted by RIS, concentrating on effective strategies to tackle real-world issues in channel estimation and beamforming design. [26] investigated the enhancement of indoor Visible Light Communication (VLC) systems through RIS technology, addressing issues related to LoS blockages and device orientation. [27] presented Intelligent Omni-Surface (IOS) for serving users in multiple directions and suggested a new hybrid beamforming approach for wireless communication based on IOS. [28] conducted a comprehensive review of IOS, discussing their application in upcoming cellular networks, design fundamentals, beamforming techniques, channel modeling, and experimental considerations. [29] examined ML algorithms for RIS, focusing on spectrum allocation in IoT systems and integrating ML techniques with their practical applications to solve technical issues. [30] reviewed recent progress in communications aided by RIS, with an emphasis on algorithms for effective CSI computation in different system setups, including massive MIMO, MISO, and cell-free configurations. [31] strengthened security in Sixth Generation (6G) networks and IoT devices, emphasizing enhanced encryption, robust authentication, and new technologies to protect data and privacy in these advanced systems. [32] investigated the integration of ML, particularly Deep Reinforcement Learning (DRL), with RIS for optimizing parameters, tackling challenges, and suggesting innovative solutions. [33] concentrated on examining the challenges in 6G networks, exploring UAV and RIS technologies, assessing existing research, and discussing their potential future amalgamation and research opportunities. [34] concentrated on a comprehensive analysis of Simultaneously Transmitting and Reflecting RIS (STAR-RIS) technology, contrasting it with conventional RIS, exploring its applications in 6G networks, categorizing various approaches, and providing perspectives on upcoming research and developments.

Additionally, existing surveys have reviewed key developments in resource management, emphasizing the use of RIS to enhance network performance in next-generation wireless networks. For example, this survey [35] explored integrating MIMO with RIS in wireless communications, enhancing performance through environmental manipulation. It examined the benefits of combining these technologies, focusing on improved resource allocation and EE via signal manipulation and beamforming. The paper analyzed key research areas, including optimization and beamforming strategies, and outlined challenges and future trends in MIMO-enabled RIS systems, highlighting their potential to transform wireless communication. In another work [36] surveyed optimization techniques for RIS-aided wireless communications in 6G networks. It outlined model-based algorithms, heuristic methods, and ML approaches, including Reinforcement Learning (RL) and Federated Learning (FL), focusing on their applications to enhance network performance. The methods were compared for stability, robustness, and optimality, emphasizing their role in improving network capacity and efficiency with low energy and cost. In [37], the overview on RISs discussed resource management, focusing on performance optimization in RIS-enhanced WNs. It detailed the design and implementation of beamforming strategies to improve network performance. Additionally, it covered the integration of ML techniques to address dynamic challenges, such as fluctuating wireless channels and user mobility, which are crucial for effective resource management and data delivery [38]. The survey also highlighted the need for ongoing research to further integrate RISs with other emerging technologies to optimize resource allocation in next-generation networks. Similarly, survey [39] discussed using RISs to enhance WNs performance, detailing their structure and operation. It covered the principles, hardware architecture, and control mechanisms of RISs, and analyzed their performance. Despite potential benefits, challenges in efficient integration and channel estimation for passive components were noted, along with comparisons of RIS deployments for different user scenarios. In another paper [40], resource management in RIS-assisted networks is explored, focusing on leveraging AI techniques. The research addresses AI-based approaches for channel estimation, phase-shift optimization, and resource allocation, essential for enhancing system performance in 6G wireless networks. These AI techniques are critical in managing the complexity of RIS-assisted systems, aiming to optimize network efficiency and effectiveness through smart resource management strategies. In [41], the authors discussed using RISs and RL to optimize wireless communication systems. It highlighted RISs as relay technology enhancements that improve signal quality, EE, and power allocation. The role of RL, especially DRL, was emphasized for optimizing RIS parameters in real-time, enhancing system performance and efficiency. Challenges and potential solutions for implementing RL in RIS technology were also noted. Additionally, [42] discussed integrating 6G IoT networks with RISs and ML to enhance system performance. It outlined RIS advancements and ML applications, including DRL and FL techniques, used to design the radio environment without pilot signals or channel state information. Additionally, it surveyed ML solutions for dynamic challenges like user movement and channel variations. The paper concluded with challenges and future directions for further integrating RISs in IoT networks.

### B. Motivation and Contribution

The impetus for this survey stems from the transformative impact of RIS in contemporary network technology. With networks becoming increasingly complex and the demand for optimized resource allocation growing, RIS stands out as a pivotal innovation. Its unique ability to dynamically manage electromagnetic wave interactions presents a compelling solution for enhancing network efficiency. A primary motivation is to explore the versatility of RIS across various network structures. From established SIMO, MISO, and MIMO frameworks to the more complex HetNets, the survey aims to assess the adaptability and efficacy of RIS. This exploration investigates RIS's potential in cutting-edge communication areas such as THz networks, VC networks, and UAV-based networks. The survey is also driven by the potential fusion of RIS with advanced technologies like AI. This exploration aims to uncover how AI can enhance the capabilities of RIS in network resource allocation, marking a new frontier in network optimization strategies. To fill this gap, we conduct a complete and in-depth study of RIS-driven resource allocation strategies for diverse network environments.

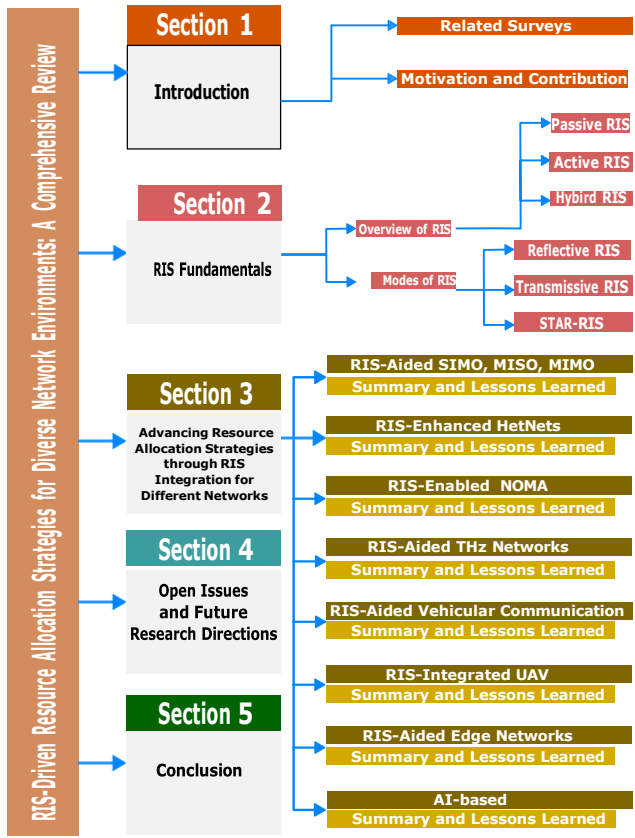

Fig. 1: Taxonomy of Resource Allocation Strategies for RIS-Driven Diverse Network Environments.

The following are the main contributions of our survey:

· This comprehensive review delves into the diverse spectrum of RIS technologies, providing an in-depth analysis of their types and operational modes. We examine active, passive, and hybrid RIS configurations, as well as their unique functionalities like reflection, refraction, and the capabilities of STAR-RIS. Our focus is on linking the theoretical underpinnings of these technologies with tangible applications, underscoring how RIS can be effectively utilized to elevate network performance in practical scenarios.

· In this survey, we methodically classify and assess various resource allocation strategies within the realm of RIS applications, spanning multiple network architectures. Our analysis offers a detailed and insightful perspective on the integration of RIS within both established and developing network infrastructures. This methodical framework aids in demystifying the complexities of RIS implementation and underscores its significant potential in optimizing network resources.

· Featuring extensive summary tables and comparative analyses, the survey offers a detailed examination of different RIS resource allocation strategies. This approach is crucial in evaluating various factors such as system models, types of RIS, CSI, and optimization techniques, thereby providing a holistic view of RIS applications in network systems.

· The survey concludes with lessons learned and a forward-looking discussion, pinpointing future research avenues and outlining the challenges that lie ahead in the realm of RIS-driven resource allocation.

This survey is meticulously structured, providing a comprehensive exploration of resource allocation within diverse network environments. Section II elucidates the fundamentals of RIS, offering a detailed overview of the types and operational modes of RIS, including passive, active, and hybrid RIS, as well as reflective, transmissive, and STAR-RIS configurations. Progressing to Section III, the survey delves into the advanced strategies for integrating RIS into various network architectures, examining how these technologies enhance resource allocation across different communication paradigms. Section IV uncovers the lessons learned, open issues, and challenges RIS faces within specialized contexts like THz communications, VC, and UAV communications, and provides an outline of potential future directions in wireless communications. The investigation reaches its culmination in Section V, where the survey draws conclusions and suggests avenues for future research, addressing the evolving landscape of network technologies and the pivotal role of RIS in shaping next-generation wireless communications. Figure. 1 illustrates the organization of the survey.

## II. RIS FUNDAMENTALS

In this section, provide an overview of RIS, a breakthrough in wireless communication. RIS includes types like passive, active, and hybrid, actively manipulating electromagnetic waves to enhance signal propagation and network efficiency. Programmable metasurfaces in RIS offer precise control over waves. RIS signal quality in environments with obstacles by redirecting and focusing signals towards intended receivers.

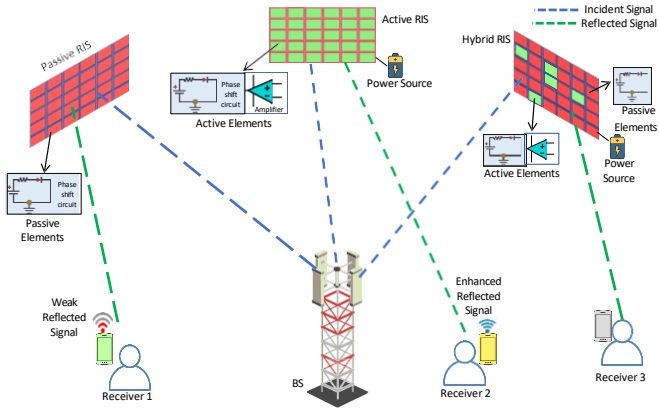

Fig. 2: Depiction of RIS types, which include passive, active, and hybrid.

Overall, RIS types provide diverse solutions for better WNs, and RIS technology introduces dual transmitting and receiving modes, improving network adaptability and efficiency.

### A. Overview of RIS

RIS technology represents a significant leap forward in wireless communication, boasting surfaces embedded with countless miniature elements. These elements wield the power to manipulate electromagnetic waves, thereby enhancing signal propagation. RIS is classified into three primary types: passive RIS, active RIS, and hybrid RIS, as shown in Figure. 2. Among their central functionalities are signal reflection, refraction, and the innovative STAR-RIS approach [43]. RIS, as advanced technology, dynamically manipulates electromagnetic waves to enhance wireless communications. Comprising numerous small elements, RIS can adjust the phase and amplitude of waves, thereby improving signal direction, strength, and interference control. These technologies find application across various domains, ranging from network efficiency enhancement to secure communications. Each offering distinct advantages such as improved network performance, lower power consumption, and enhanced signal coverage.

Despite the transformative potential of RIS technology in WNs, challenges persist. These challenges include intricate hardware design, seamless integration with existing network infrastructures, and the development of sophisticated control algorithms. Researchers are actively addressing these obstacles, aiming to fully harness RIS capabilities for the advancement of efficient wireless communication systems [34].

*1) Passive RIS:* Passive RIS cost-effective materials in various applications, such as building facades and wearable technology, to bolster wireless communication, as shown in Figure. 2. Unlike traditional relay systems like Amplify-and-Forward and Decode-and-Forward, RIS operates without the need for heavy power requirements. It functions by passively reflecting and phase-shifting incoming signals, enabling environmentally friendly and efficient deployment. However, despite its ability to support full duplex and full-band transmission solely through signal reflection, passive RIS technology is susceptible to a phenomenon known as "Double Fading." This phenomenon results in significant signal degradation, reducing the overall efficiency of the system compared to direct links. To address this limitation, active RIS designs have been developed, aimed at mitigating multiplicative fading and enhancing overall performance [44], [45].

*2) Active RIS:* As shown in Figure. 2, active RIS improves upon passive RIS technology by actively amplifying and adjusting the phase of reflected signals. Utilizing phase shift circuits and amplifiers, active RIS systems consume more power compared to their passive counterparts. However, this active approach allows for the conversion of multiplicative channel fading into additive ones, thereby enhancing overall system effectiveness. Despite concerns regarding increased energy consumption, active RIS holds promise for delivering cost-effective and efficient solutions when compared to traditional relay systems [46]–[48].

*3) Hybrid RIS:* Hybrid RIS combines active and passive elements to enhance wireless communication, as shown in Figure. 2. It supports simultaneous reflection and transmission of signals for 360-degree coverage, addressing traditional RIS limitations. Hybrid RIS operates as a 'single-input, dual-output' system, considering energy, mode, and timing in its protocol. Despite its complexity and potential high cost, it offers improved adaptability and effectiveness, potentially including comprehensive indoor and outdoor coverage [49].

### B. Modes of RIS

RIS wireless communications by manipulating electromagnetic waves. They operate in three primary modes: Reflective, Transmissive, and STAR. Reflective RIS modifies signals back towards the transmitter, ideal for direct LoS scenarios. Transmissive RIS allows signals to penetrate, reaching users behind obstacles. STAR-RIS combines both functionalities, managing signals in multiple directions for complex environments. These modes improve network adaptability and efficiency, addressing diverse communication demands [50]. Figure. 3 illustrates three operational modes of RIS: Reflective RIS, Transmissive RIS, and STAR-RIS.

*1) Reflective RIS:* As shown Figure. 3, a reflecting RIS is an engineered metasurface composed of multiple small meta-atoms, which can manipulate electromagnetic waves in a controlled manner. In the context described, the RIS employs varactor diodes to adjust the biasing voltages, thus altering the reflection coefficients of the meta-atoms. This modification enables the RIS to control both the amplitude and phase of harmonics generated when illuminated by electromagnetic waves. The time-domain digital coding technique applied allows for precise control over these parameters, facilitating advanced wavefront shaping for various harmonic frequencies. This technology is primarily demonstrated at microwave frequencies but has potential applications in the THz and optical regimes. Additionally, it can be adapted to transmission type configurations for further control over harmonic manipulation using an array of patch antennas [50], [51].

*2) Transmissive RIS:* A transmissive RIS is a type of metasurface that allows electromagnetic signals to penetrate through it, thereby serving users located on the opposite side of the BS. This capability is particularly useful in scenarios where direct LoS communication is obstructed, as the transmissive RIS can modify the phase, amplitude, and potentially other properties of the transmitted waves to ensure effective signal propagation and coverage to areas beyond the physical location of the RIS. This implementation is distinct from the reflective type, which redirects signals back towards the same side of the BS, and the hybrid type, which combines both reflective and transmissive functionalities to cater to a broader range of communication scenarios [51], [52].

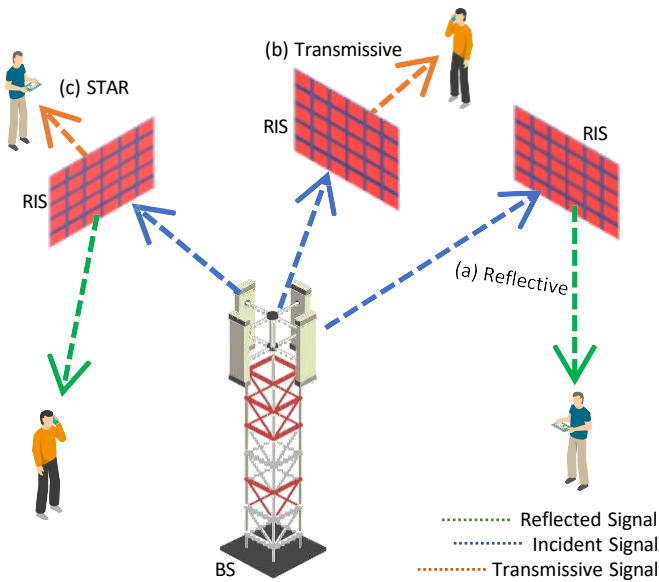

Fig. 3: RIS operational modes that include reflective, transmissive, and STAR.

*3) STAR-RIS:* The STAR-RIS is an advanced form of traditional RIS that enhances signal coverage and adaptability by managing both reflection and transmission of wireless signals [43]. Unlike conventional RIS systems, which only reflect signals and thus only interact with users on the same side of the BS, the STAR-RIS can handle users located on both sides of the surface. As shown in Figure. 3. It achieves this by splitting incoming signals into two distinct components: one part is reflected back, and the other is transmitted through the surface, providing 360-degree coverage. This dual-functionality makes STAR-RIS particularly valuable in complex environments where signals need to be managed over multiple spatial regions, such as from one room to another or from outside to inside a building. However, the design and implementation of STAR-RIS involve challenges related to energy consumption, system modes, and hardware complexity, which are essential to address to fully exploit its potential for enhancing wireless communication coverage comprehensively [34].

The simultaneous presence of these two modes within RIS technology marks a significant advancement in the wireless communication domain. It grants networks the versatility to effortlessly switch between enhancing signal transmission and conducting detailed analysis of incoming signals, consequently improving both the network's efficacy and its adaptability. This dual functionality establishes RIS as a crucial component in the future of wireless systems, equipping them to tackle the intricate and ever-changing demands of modern and future communication requirements. The dual-mode capability not only highlights the adaptability of RIS but also lays the groundwork for smarter, more effective, and reactive WNs, signifying a pivotal development in the progression of wireless communication technology.

## III. ADVANCING RESOURCE ALLOCATION STRATEGIES THROUGH RIS INTEGRATION FOR DIFFERENT NETWORKS

In this section, our primary aim is to investigate how RIS to the optimization of resource allocation within network systems. We methodically classify and assess the applications of RIS in various network architectures, including SIMO, MISO, and MIMO systems. Our analysis extends to the incorporation of RIS in complex network settings, such as HetNets, and its utilization within NOMA frameworks. Furthermore, we explore the implications of RIS in advanced communication domains, such as THz networks, VC, and UAV communications, with a particular emphasis on how RIS collaborates with AI to enhance overall network efficiency.

### A. RIS-Aided Resource Allocation for SIMO, MISO, and MIMO Systems

This subsection examines RIS-aided resource allocation within SIMO, MISO, and MIMO systems, with a focus on optimizing key design variables such as power allocation, beamforming, and phase shifts. In SIMO systems, RIS integration has led to innovative solutions that maximize SINR and improve EE, especially with multiple RIS configurations. For MISO systems, optimized beamforming and power allocation strategies have significantly enhanced multi-user capacity and overall system performance. In MIMO systems, RIS plays a critical role in overcoming challenges like link blockage and maintaining information freshness by optimizing power, precoding, and RIS phase shifts, as shown in Figure. 4. The design variables highlighted in the summary table—such as transmit power, RIS coefficients, and beamforming vectors—demonstrate the effectiveness of various optimization techniques in boosting EE, maximizing WSR, and improving system capacity across these communication models. This cohesive approach showcases how RIS technology is transforming resource allocation strategies to meet the evolving demands of modern wireless networks.

The paper [53] introduced a simple projected gradient descent-based solution for optimizing RIS phases in next-generation communication systems. By leveraging channel statistics instead of instantaneous CSI, the algorithm maximizes the minimum Signal-to-Interference and Noise Ratio (SINR) and ensures fair operation among users. The approach significantly reduces computational complexity and control data exchange, making it practical for implementation. The algorithm also caters to electromagnetic fields (EMF)-aware systems. Numerical results highlight its superiority compared to systems with random RIS phase shifts. In another work [54], the authors explored uplink transmission in a multiuser SIMO system assisted by multiple RISs, focusing on maximizing EE

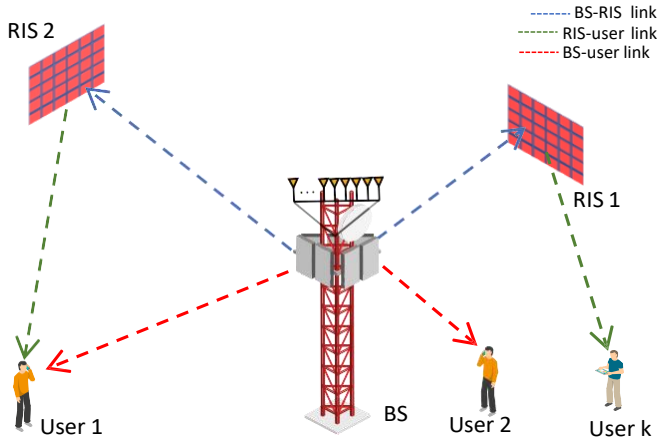

RIS 2

RIS 1

BS

User 1   BS   User 2   User k

······ BS-RIS link
······ RIS-user link
······ BS-user link

Fig. 4: Depiction of multi-RIS-aided downlink MU-MIMO network.

with EMF exposure constraints. It introduced the Energy Efficient Multi-RIS (EEMR) algorithm to optimize user transmit power and RIS phase shifts. Comparisons showed that systems with distributed RISs were more energy-efficient than those with a central RIS, highlighting the trade-off between EE and EMF awareness due to exposure limits.

Different from Orthogonal Multiple Access (OMA) schemes [53], [54], in this paper [55], NOMA system-based Physical Layer Security (PLS) of an active RIS-assisted uplink SIMO was explored. An optimization problem was formulated to maximize the System Sum Rate (SSR) by jointly designing the BS's receive beamforming vector, active RIS reflecting coefficients, and user transmit power. The Block Coordinate Descent (BCD) method was applied to solve the problem iteratively. Simulation results demonstrated that the active RIS-assisted scheme surpassed passive RIS-based and OMA-based schemes in SSR performance.

The authors focused on energy-efficient resource allocation in RIS-assisted MISO communication systems. In [56], the authors discussed leveraging RIS in wireless communication to enhance multi-user capacity and meet QoS requirements. The authors introduced a beamforming optimization strategy at the BS utilizing Zero-Forcing (ZF) beamforming and power distribution methods. They calculated the optimal power distribution using mathematical techniques and refined the RIS phase shift strategy through a gradient descent method. Simulation outcomes indicate that this optimization approach significantly enhances capacity and surpasses random reflecting phase shift strategies. In another paper [57], they introduced two different algorithms (the Wolfe-based Gradient-descent (GAW) EE maximization algorithm and the Trust Region (TR)-based EE maximization algorithm) for enhancing EE and apply Dinkelbach's algorithm for optimal power distribution. The outcomes demonstrated that these new methods surpass current techniques, underlining the importance of their contributions to boosting EE in networks supported by RIS.

This study [58] proposed EE maximization in WNs using RIS, applying a unified reflection constraint to all RIS elements to expand feasible solutions and enhance performance. It introduced two polynomial complexity algorithms to optimize RIS coefficients, users' transmit powers, and BS filters. Theoretical analyses and numerical results assessed the effectiveness of these methods, determining when active or nearly passive RISs are preferable for EE. Similarly, in this paper [59], the authors put forward a robust optimization framework for RIS-assisted UAV MISO communication networks, addressing UAV's battery limitations and oscillations. It focused on maximizing the WSR by optimizing BS beamforming and RIS's phase shifts under UAV oscillations. The proposed method combined BCD and Lagrangian dual transform, showing superior performance in simulations compared to benchmarks.

Contrasting the aforementioned works [56]–[59] that are leveraging single RIS, the authors in [60] introduced a framework with both fixed and mobile RISs. They proposed a WSR maximization problem in a cooperative RIS-assisted downlink MISO system. It tackled the WSR maximization by alternating active and dual-RIS passive beamforming, utilizing the Weighted Minimum Mean Square Error (WMMSE) and the Sequence Phase-Rotation (SPR) or the Semi-Definite Relaxation (SDR). Simulations indicated that the SPR method outperformed benchmarks, with better WSR performance in the nearfield than in the far-field.

Unlike the above works considering SIMO and MISO-based systems, this paper [61] considered a MIMO communication system and proposed a method to optimize the precoder and multiple RISs in a downlink setting. By partitioning the RISs into tiles and solving a tile assignment problem, the complexity of RIS optimization is reduced. The objective is to achieve fairness among users while minimizing the WSR. Simulation results highlight the potential of RISs to significantly benefit users with severe path losses when appropriately designed. In another work [62], the authors considered MU-MIMO and investigated the downlink scenario with multiple RISs. They introduced two Alternating Optimization (AO) frameworks to tackle the non-convex resource allocation challenge, focusing on optimizing beamforming at both the BS and RISs. The minimum mean square error (MMSE) and Riemannian conjugate gradient (RCG) approach demonstrates fast convergence and enhanced EE, while the sequential fractional programming (SFP) approach exhibits improved spectral performance. Both techniques outperform random phase-shift settings in terms of performance improvement.

This article [63] addressed RIS-assisted mmWave communications to mitigate link blockage and enhance information freshness. The study focused on maximizing SSR and satisfying information freshness by optimizing beamforming, RIS reflection coefficients, and UE scheduling. The problem was divided into subproblems of data rate optimization and scheduling strategy design, solved using hierarchical and local search methods. Simulations showed the algorithm effectively improved the SSR while meeting information freshness requirements. The authors in [64] discussed a double-RIS aided downlink MIMO system, focusing on the mean-square-error (MSE) problem minimization by optimizing transmit beamforming, receive equalizer, and RIS passive beamforming. It introduced a common reflection pattern for both RISs to simplify complexity. A majorization-minimization based algorithm was used to address the tighter variable coupling.

Numerical results showed that this approach nearly matched the performance of separate reflection patterns while reducing complexity.

A transmission framework for mmWave MIMO systems using a RIS was proposed [65]. It uses partial CSI to analyze the ergodic mutual information (EMI), which stabilizes at a constant in high-Signal-to-Noise Ratio (SNR) scenarios. An algorithm was developed to optimize RIS and BS beamforming, enhancing array gain and EMI. Numerical results confirmed these findings. In another paper [66], the authors considered imperfect CSI and investigated integrating a RIS into a secure multiuser massive MIMO system, addressing hardware impairments, and proposed a power allocation strategy to optimize Secrecy Rates, confirmed through simulations. Different from [65], [66], which considered partial and imperfect CSIs, respectively, this paper [67] assumed perfect CSI and proposed rate maximization techniques for single-user and MU-MIMO systems using the WMMSE criterion, optimizing RIS phase shifts to maximize sum-rate and minimize weighted mean square error (WMSE). By employing a tensor-based RIS system model, it detailed strategies for phase shift optimization, deriving the closed-form gradient of WMSE and sum-rate with respect to the RIS phase shift vector. Simulations revealed that the proposed WMMSE-based rate maximization technique outperformed other benchmarks.

*Summary and Lessons Learned:* **Summary:** Recent studies on RIS-aided resource allocation in SIMO, MISO, and MIMO systems have primarily focused on leveraging perfect and statistical CSI to optimize various network parameters. For single RIS setups, perfect CSI has been predominantly used to optimize beamforming, power, and phase shifts in SIMO systems [53], and MISO systems have explored EE optimization using algorithms such as gradient descent and fractional programming [57]. UAV-based MISO networks also optimize using robust schemes like the block coordinate Lagrangian dual transform under known CSI [59]. However, the critical gap lies in the scalability of these algorithms when extended to larger, more complex networks, especially under dynamic conditions.

In multi-RIS deployments, known CSI has been used to optimize power and phase shifts in MU-SIMO systems [54]. However, the complexity of managing multiple RIS units introduces significant computational challenges, which are not adequately addressed in the literature. For cooperative RIS-assisted MISO systems, WSR maximization is achieved through alternating active and passive beamforming under known CSI [60]. However, there is a notable lack of research on the impact of imperfect CSI in these multi-RIS environments. Although the EEMR algorithm shows promise in improving EE [54], its applicability in real-world scenarios with dynamic traffic patterns and varying channel conditions remains uncertain.

These studies, as summarized in Table II, highlight the advancements made in RIS-aided resource allocation for SIMO, MISO, and MIMO systems, but also reveal significant gaps. Key challenges include developing more scalable multi-RIS deployment strategies, effectively handling imperfect CSI, and creating algorithms that can adapt to real-time changes in network conditions. Addressing these gaps is crucial for the broader adoption and effectiveness of RIS technology in diverse network environments.

**Lessons Learned:** The deployment and optimization of RIS in SIMO, MISO, and MIMO systems play a crucial role in enhancing EE and spectral efficiency. The performance of these systems is influenced by several factors, including the number of RIS elements, phase shift strategies, power management, and hardware design. Below are key insights gained from effectively deploying RIS in these systems:

· RIS Size: To maximize EE in both centralized and distributed RIS-aided SIMO systems, it's important to determine the optimal number of RIS elements. Centralized systems often perform better with a smaller number of elements, whereas distributed systems benefit from a larger number. Continuous phase shifts can provide higher total rates, but discrete phase shifts are generally more energy-efficient due to lower internal power consumption. Therefore, the balance between continuous and discrete phase shifts, along with the right number of RIS elements, is crucial for optimizing system efficiency [54].

· Choosing the Right RIS Configuration: For both single-user and multi-user systems, RIS configurations with global reflection capabilities are more effective in maximizing EE. Active RISs tend to outperform nearly-passive ones, provided that the extra power consumption remains manageable. This emphasizes the need for careful selection of RIS configurations that optimize power consumption while delivering high system performance [58].

· Balancing RIS Elements and EE: Adding more RIS elements can improve both spectral efficiency and EE by increasing the diversity of channel gains. However, beyond a certain threshold, the increased static power consumption can outweigh these benefits, leading to a decrease in overall EE. It is crucial to find the right balance between the number of RIS elements and energy consumption to ensure optimal system performance [62].

· Optimizing Power Allocation: Increasing the transmit power in multi-RIS-aided MIMO systems can improve spectral efficiency, but its impact on EE is limited, as the gain in sum-rate does not scale proportionally with the power increase. Allocating power at a more granular level, such as by clusters, allows for more precise control, which can lead to better resource allocation and user fairness. In scenarios with uniform channel conditions, a centralized power budget at the BS can simplify the power allocation process [62].

· Adapting to Varying User Densities: A rise in user density can enhance both spectral and EE, but it requires careful system adjustments. Researchers should tailor resource allocation strategies, transmission power, and the number of reflecting elements to the specific user density and distribution patterns in different environments. By doing so, the system can maintain high efficiency and performance across a range of conditions [62].

· Ensuring Robust Hardware Design: Robust hardware de-

TABLE II: Summary of Resource Allocation Schemes for SIMO, MISO and MIMO

| Ref | Year | System Model | No's | A | P | H | R | T | STAR | CSI | Designed Variable | Optimization Methodology |
|---|---|---|---|---|---|---|---|---|---|---|---|---|
| | | | | **Types** | | | **Modes** | | | | **Design and Optimization** | |
| [53] | 2023 | UL, SIMO, BSs (MA), UEs (SA) | Single | ✗ | ✓ | ✗ | ✓ | ✗ | ✗ | Statistical | Beamforming, power and phase shifts | A simple projected gradient descent-based solution |
| [54] | 2024 | UL, MU-SIMO, BS (MA), UEs (SA) | Multiple | ✗ | ✓ | ✗ | ✓ | ✗ | ✗ | Known | Transmit power and phase shifts | The Energy Efficient Multi-RIS (EEMR) algorithm |
| [55] | 2024 | UL, SIMO-NOMA, BS (MA), UEs (SA), Eves (SA) | Single | ✗ | ✓ | ✗ | ✓ | ✗ | ✗ | Known | Beamforming, reflecting coefficients, transmit power | BCD, SCA |
| [56] | 2023 | DL, MISO, BS (MA), UEs (SA) | Single | ✓ | ✗ | ✗ | ✓ | ✗ | ✗ | Known | Beamforming, power and phase shifts | Gradient descent |
| [57] | 2023 | DL, MISO, BS (MA), UEs (SA) | Single | ✓ | ✗ | ✗ | ✓ | ✗ | ✗ | Known | Powers and phase shifts | GAW, TR and Dinkelbach |
| [58] | 2024 | UL, MISO, BS (MA), UEs (SA) | Single | ✓ | ✓ | ✗ | ✓ | ✗ | ✗ | Known | Reflection coefficients of RIS, UE's transmit power and linear filters at the BS | Two provably convergent EE maximization algorithms |
| [59] | 2024 | DL, MISO, BS (MA), UEs (SA), UAV | Single | ✗ | ✗ | ✗ | ✓ | ✗ | ✗ | Known | Transmit beamforming of BS and the phase shift of RIS | A robust block coordinate Lagrangian dual transform (RoBI) scheme by leveraging the BCD and Lagrangian dual transform |
| [60] | 2024 | DL, MISO, BS (MA), UEs (SA) | Single | ✗ | ✓ | ✗ | ✓ | ✗ | ✗ | Known | The active beamforming and the dual-RIS passive beamforming | MMSE, SPR and SDR |
| [61] | 2023 | DL, MIMO, BS (MA), UEs (MA) | Multiple | ✗ | ✓ | ✗ | ✗ | ✗ | ✗ | Known | Power, precoding coefficients and phase-shifts | WMMSE |
| [62] | 2023 | DL, MU-MIMO, BS (MA), UEs (SA) | Single | ✗ | ✓ | ✗ | ✓ | ✗ | ✗ | Known | Beamforming and power | MMSE, RCG and SCA |
| [63] | 2024 | DL, MIMO, BS (MA), UEs (MA) | Single | ✗ | ✗ | ✗ | ✓ | ✗ | ✗ | Partial | The beamforming at transceivers, the discrete RIS reflection coefficients, and the UE scheduling strategy | BCD and a low-complexity heuristic scheduling algorithm |
| [64] | 2024 | DL, MU-MIMO,BS (MA), UEs (MA) | Multiple | ✓ | ✗ | ✗ | ✓ | ✗ | ✗ | Perfect | The active transmit beamforming, the receive equalizer and the passive beamforming at each RIS | A majorization-minimization (MM) based AO algorithm |
| [65] | 2024 | DL, MIMO, BS (MA), UT (MA) | Single | ✗ | ✓ | ✗ | ✓ | ✗ | ✗ | Partial | The passive beamformer of RIS and the active beamformer of BS | An efficient joint passive and active beamforming algorithm |
| [66] | 2024 | DL, MISO, BS (MA), UEs (SA), Eve (SA) | Single | ✗ | ✓ | ✗ | ✓ | ✗ | ✗ | Imperfect | Power | A power allocation strategy |
| [67] | 2024 | DL, MU-MIMO, BS (MA), UEs (SA) | Single | ✗ | ✓ | ✗ | ✓ | ✗ | ✗ | Perfect | RIS phase shift | WMMSE |

*DL-Downlink, UL-Uplink, SA-Single Antenna, MA-Mutiple Antennas, MU-Mutiple Uers, UT-User Terminals*

sign is essential to fully realize the benefits of RIS in sophisticated communication systems, particularly under imperfect CSI conditions. Although RIS can maintain high secrecy rates with an increasing number of elements, hardware impairments can significantly reduce these benefits. This underscores the importance of improving channel estimation methods and developing resilient hardware designs to enhance the security and effectiveness of RIS in advanced communication settings [66].

### B. RIS-Enhanced Resource Allocation in Heterogeneous Wireless Networks

This subsection examines resource allocation strategies in RIS-enhanced HetNets, as shown in Figure.5, focusing on the optimization of key design variables, including power allocation, beamforming, and phase shifts. The studies reviewed highlight the role of RIS in enhancing energy efficiency, reducing latency, and improving overall system performance in future 6G networks. Techniques such as AO, BCD, and SCA have been utilized to manage these variables effectively across various network scenarios. In the majority of the works, single RIS configurations are explored, where optimizing transmit power and RIS phase shifts shown to significantly enhance system performance. For instance, optimizing power allocation can maximize secrecy rates, while fine-tuning beamforming and phase shifts can lead to improved energy efficiency. Notably, there is one study that considers multiple RIS configurations, which further optimizes sub-carrier assignment, user association, power allocation, and phase shifts, demonstrating potential improvements in energy efficiency and throughput in densely populated network environments.

The summary Table III given at the end underscores the diversity of system models and the complexity of the optimization techniques employed, particularly in single RIS setups. Despite these advancements, optimizing these design variables in dynamic and large-scale HetNets remains challenging, signaling the need for more scalable and adaptive algorithms that can respond to real-time network conditions. Continued research is vital for the successful deployment of RIS in increasingly heterogeneous and evolving network landscapes.

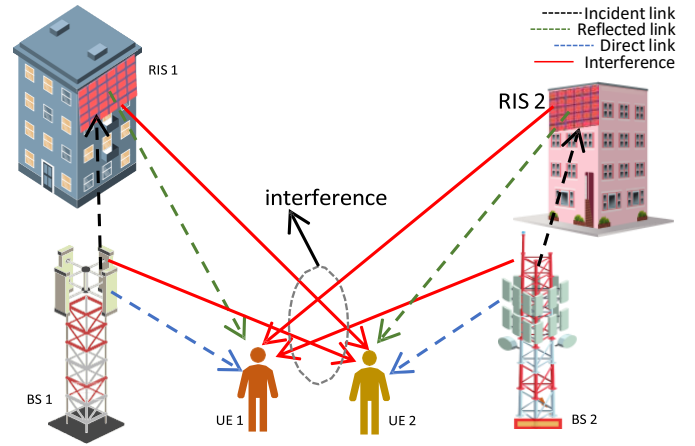

Fig. 5: illustration of multiple RIS-assisted HetNets.

In [68], introduced a framework that enhances the effective EE by optimizing transmit powers and RIS phase shifts in a multicarrier link. The numerical data confirmed the effectiveness of this approach in boosting system performance. The authors concentrated on evaluating the secrecy rate in RIS-enhanced WNs with an eavesdropper present. In [69], they suggested a power allocation strategy and calculated the optimal power distribution coefficient to maximize the secrecy rate. Their findings emphasized the superiority of the system model that includes a controlling jammer and illustrated the benefits of optimal power allocation compared to uniform power distribution. This paper [70] proposed a Dinkelbach and majorization-minimization (DMM) aided joint passive beamforming and power allocation optimization (JPB-PAO) scheme for RIS-assisted wireless systems, considering pilot-embedded data transmission and a transmit energy constraint. The DMM-JPB-PAO scheme is designed to optimize pilot power, data power, and RIS phase shifts. Results demonstrated its effectiveness compared to baseline schemes, achieving high achievable rates with reduced computational complexity. In another paper [71], the authors tackled the resource allocation issue in an RIS-supported heterogeneous network. The authors developed both centralized and distributed algorithms to optimize sub-channel allocation, transmit power, and RIS coefficients. These algorithms were assessed and compared using numerical analysis, highlighting the benefits of each method.

Unlike the OMA-based uplink and downlink scenario works [69]–[71], the authors in [72] studied uplink NOMA-based RIS system for future cellular networks, focusing on Sparse Code Multiple Access (SCMA) compatibility. They optimized an RIS-assisted network to maximize sum-rate, using a BCD method. Their results showed significant improvements over conventional methods. Similarly, in [73], the authors introduced a wireless-powered NOMA system with STAR-RIS, analyzing its performance by outage probability, throughput, and Age of Information (AoI). Their strategies, time-switching and energy-splitting policy (TEP) and double energy-splitting policy (EEP), improved throughput and maintained acceptable outage and AoI levels. They also developed the genetic-algorithm based time allocation and power allocation (GA-TAPA) algorithm to optimize throughput while meeting AoI criteria.

This study [74] investigated a system that combines STAR-RIS with full-duplex relay technology to maximize downlink sum-rate. They optimized beamforming at the relay, BS, and STAR-RIS using a penalty-based algorithm with Successive Convex Approximation (SCA). Numerical results showed significant improvements over baseline methods. Similarly, the authors in [75] explored traffic offloading in HetNets using Device-to-Device (D2D) communications and RIS to enhance spectrum efficiency. They addressed resource allocation through game theory, optimization, and local search methods, with simulations showing their approach outperformed existing benchmarks.

Different from the works [68]–[75] that considered single RIS, in this work [76], the authors assumed multiple RISs into HetNets to minimize energy consumption in densely deployed small cells. They tackled this with a multi-objective optimization, using an enhanced non-dominated sorting genetic algorithm II (NSGA-II) algorithm to optimize throughput and EE. Numerical results showed that optimizing RIS deployment significantly improves network performance. In another work [77], the authors considered a RIS-assisted HetNets model for multi-base, multi-frequency networks. The authors aimed to maximize the SSR by addressing user resource allocation and RIS phase shift optimization using the BCD method and a local discrete phase search algorithm. Simulations showed this approach improved the SSR and achieved near-optimal performance with low complexity. In another paper [78], the authors investigated a RIS-aided MEC HetNets to minimize total latency through optimized service caching, task offloading, and resource allocation. The NP-hard optimization was split into communication and computing sub-problems, effectively reducing the total delay, as shown by numerical results.

*Summary and Lessons Learned:* **Summary:** Recent advancements in RIS-aided resource allocation for HetNets, as detailed in Table III, emphasize the optimization of power allocation, beamforming, and phase shift adjustments. These studies predominantly focus on single RIS setups, leveraging known CSI to enhance network efficiency and capacity. For example, in OFDM systems, transmit powers and RIS phase shifts are optimized using AO and fractional programming [68]. In secure downlink systems with jammers, power allocation strategies are applied to improve performance under known CSI conditions [69]. Uplink systems involving APs and UEs employ passive beamforming and phase shift optimization using a quasi-static approach [70], while macro BS systems focus on sub-channel allocation and RIS coefficients using AO [71].

In more complex scenarios, such as SCMA-based uplink systems with BS, UEs, and D2D pairs, iterative BCD algorithms effectively manage power and phase shift allocation [72]. NOMA-based systems optimize time and power allocation using TEP, EEP schemes, and GA-TAPA algorithms under perfect CSI [73]. Downlink relay systems optimize beamforming using AO and SCA methods under perfect CSI [74]. Additionally, multi-RIS setups with known CSI optimize sub-carrier assignment, UE association, power allocation, and phase shifts using NSGA-II algorithms [76]. Single RIS-assisted HetNets with known CSI utilize BCD and local discrete phase search algorithms to optimize resource allocation and phase shifts [77], while MEC HetNets employ local-search and genetic algorithms to optimize passive beamforming at RIS and system bandwidth [78]. Despite these advancements, a critical gap remains in addressing scenarios with imperfect CSI, which is underrepresented across these studies. Focusing future research on imperfect CSI could significantly improve the robustness and adaptability of RIS-aided resource allocation strategies in HetNets.

**Lessons Learned:** Deploying RIS in communication systems requires careful consideration of factors such as the number of elements, phase shift optimization, power management, and overall hardware design. These factors play a critical role in determining system performance, EE, and communication quality. The following are key insights for optimizing RIS deployment:

· Phase Shift Optimization: Adjusting the phase shifts in RIS is essential for enhancing signal quality and minimizing interference within the network. By carefully optimizing these shifts, the system can improve signal reception and reduce co-channel interference among users. In high-density network environments, communication can lead to better spectrum utilization, less strain on BSs, higher transmission speeds, and an overall improved user experience by effectively managing phase shifts [75].

· Strategic Placement of RIS: Positioning RIS closer to user devices in HetNets can significantly boost system performance. However, this improvement tends to diminish beyond a certain distance, indicating that there is an optimal range for deploying RIS to achieve maximum effectiveness [76].

· Balancing BS and RIS Use: While increasing the number

TABLE III: Summary of Resource Allocation Schemes for Heterogeneous Wireless Networks

| Ref | Year | System Model | RIS Details | | | | | | | | CSI | Design and Optimization | |
|---|---|---|---|---|---|---|---|---|---|---|---|---|---|
| | | | No's | Types | | | Modes | | | | | Designed Variable | Optimization Methodology |
| | | | | A | P | H | R | T | STAR | | | | |
| [68] | 2023 | OFDM, Transmitter (SA), Receiver (SA) | Single | X | ✓ | X | ✓ | X | ✓ | | Known | Transmit powers and RIS phase shifts | AO and fractional programming algorithm |
| [69] | 2023 | DL, Source (SA), Destination (SA), Eve (SA), Jammer (SA) | Single | X | X | ✓ | ✓ | X | X | | Known | Power | Lagrange's multiplier approach, a power allocation strategy |
| [70] | 2023 | UL, AP (SA), UEs (SA) | Single | ✓ | X | X | ✓ | X | X | | Quasi-static | Passive beamforming, power and phase shift | Termed as DMM-JPB-PAO |
| [71] | 2023 | UL, Macro BS (MA), Small BSs (SA), Terminal device (SA) | Single | ✓ | X | X | ✓ | X | X | | Known | Sub-channel allocation, transmit power and RIS co-efficients | AO |
| [72] | 2023 | UL, SCMA, BS (MA), UEs (SA), D2D pairs (SA) | Single | X | ✓ | X | ✓ | X | X | | Known | Transmitted power and the phase shifts | An efficient iterative algorithm based on BCD method |
| [73] | 2023 | UL and DL, NOMA, AP (SA), UEs (SA) | Single | X | X | ✓ | ✓ | X | ✓ | | Perfect | Time and power | TEP and EEP schemes & GA-TAPA algorithm |
| [74] | 2023 | DL, Relay (MA), BS (MA), UEs | Single | X | X | ✓ | X | X | ✓ | | Perfect | Beamforming | AO and SCA |
| [75] | 2023 | UL, HetNets, BS (MA), Devices (MA) | Single | ✓ | X | X | ✓ | X | X | | Known | Power and RIS phase shift | A coalitional game method based on the game theory |
| [76] | 2023 | DL, HetNets, BSs (MA), UEs (SA) | Multiple | X | ✓ | X | ✓ | X | X | | Known | The sub-carrier assignment, UEs association, power allocation, and RISs' phase shift | NSGA-II algorithm |
| [77] | 2023 | DL, HetNets, BSs (MA), UEs (MA) | Single | X | ✓ | X | ✓ | X | X | | Known | UE resource and the phase shift of RIS | BCD and a local discrete phase search algorithm |
| [78] | 2024 | UL, MEC, HetNets, Macro cell BS (MA), Small cell BSs (SA), Macro cell UEs (SA), Small cell UEs (SA) | Single | X | ✓ | X | ✓ | X | X | | Known | Passive beamforming at RIS and system bandwidth | The local-search and genetic-algorithm |

of BS can enhance network performance, the benefits taper off after a certain point. Incorporating RIS can further improve system efficiency and reduce energy consumption, potentially minimizing the need for additional BSs. The key is to find the right balance between the number of BSs and RIS to optimize both performance and energy use [76].

· Managing High User Density: The system's performance remains stable even as the number of users increases, up to a certain limit. RIS-enhanced systems are particularly effective in maintaining high-quality communication in densely populated networks, demonstrating RIS's ability to support more users without a significant drop in performance [76].

### C. RIS-Enabled NOMA Resource Allocation Strategies

This subsection reviews resource allocation strategies for RIS-enabled NOMA systems, as depicted in Figure 6, focusing on optimizing critical design variables such as power allocation, beamforming, phase shifts, user pairing, and decoding order. The integration of STAR-RIS in NOMA systems has shown significant potential in improving key performance metrics like spectrum efficiency, system throughput, and energy efficiency. These improvements are achieved through advanced optimization techniques such as AO, BCD, and iterative methods, which systematically optimize design variables to enhance system performance. The reviewed studies highlight various approaches, including single and multiple RIS configurations, to tackle challenges in NOMA systems, particularly under conditions of imperfect or statistical CSI. These configurations aim to improve connectivity, reduce interference, and maximize system resources. The studies demonstrate how optimizing variables like transmit power, RIS configurations, and reflection coefficients can lead to substantial gains in resource utilization and overall network performance. However, the optimization of these design variables, particularly in large-scale and dynamic environments, remains complex. The need for scalable and robust strategies is evident, as managing the interactions between these variables in real-time is challenging. The summary table at the end outlines the diverse system models and optimization methodologies employed, emphasizing the importance of further research to develop more adaptive and efficient algorithms capable of handling the intricacies of RIS-enabled NOMA systems in evolving network scenarios.

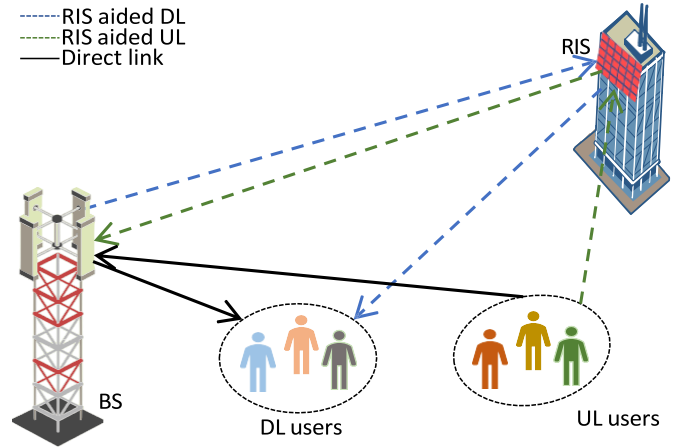

Fig. 6: Depiction of RIS-aided full duplex NOMA system.

This article [79] integrated NOMA and over-the-air federated learning (AirFL) with a STAR-RIS. It managed interference, improved coverage, and enhanced learning performance. Researchers optimized user transmit power and STAR-RIS configurations, leading to improved spectrum efficiency and faster convergence in simulations. Similarly, in [80], the authors studied a system with a STAR-RIS aiding NOMA with two users. They proposed a power and amplitude allocation scheme to improve the reflected user's service quality. The authors derived formulas for outage probability and diversity order under Nagakami-fading conditions. This approach notably enhanced the reflected user's performance and overall system throughput. In another work [81], the authors studied resource allocation in STAR-RIS-assisted multi-carrier networks to maximize the system sum-rate. For both orthogonal and NOMA, they developed optimization methods. STAR-RIS-aided NOMA networks outperformed conventional RISs and OMA, with effective strategies identified for both.

Unlike the perfect CSI assumption in the works [79]–[81], the authors in this article [82] assumed statistical CSI and studied the rate performance of a STAR-RIS-aided NOMA system, optimizing power allocation and amplitude coeffi-cients for Effective Rate (ER) maximization. In another paper [83], RIS utilization in NOMA for enhancing spectral-efficient Full-Duplex (FD) Ultra-Reliable Low-Latency Communication (URLLC) systems. On maximizing sum-rate with power allocation/beamforming optimization. Numerical results showed orthogonal and Half-Duplex (HD) schemes, with RIS enhancing rate performance. The authors in [84] explored resource allocation in RIS-aided FD NOMA networks. An AO-based approach for joint parameter optimization, outperforming traditional methods and achieving comparable results to exhaustive search. Results underscored the method's effectiveness in improving in RIS-aided networks. Similarly, the authors in [85] proposed a novel system integrating STAR-RIS with MEC to minimize user energy consumption. A decomposition approach to solve the non-convex optimization problem. Simulations indicated that STAR-RIS deployment encouraged more task bit offloading and outperformed traditional RISs in sum energy consumption, particularly for larger task-input bits.

Different from the works in [79]–[85], which considered single RIS, the authors in [86] introduced a double RIS-aided ambient backscatter communication network integrated with NOMA technology to enhance connectivity and improve various performance metrics. Addressed the resource allocation issue by optimizing phase shifts and power allocation. Simulation results confirmed the effectiveness of their proposed approach.

This article [87] addressed EE optimization in RIS-aided mmWave networks with NOMA, focusing on joint resource allocation for power and beamforming. Using iterative algorithms integrating majorization-minimization and BCD, the study derived suboptimal solutions for beamforming and power allocation. Simulation results demonstrated that this approach achieved higher EE with lower complexity compared to benchmark schemes. In another work [88], the authors explored a hybrid NOMA framework with STAR-RIS to enhance cell-edge communication, focusing on maximizing the minimum downlink rate by optimizing user pairing, decoding order, beamforming, and resource allocation. A two-layer iterative algorithm was introduced to tackle the complex problem, with simulations showing significant performance improvements over traditional methods and emphasizing the role of beamforming and power allocation.

This paper [89] examined RIS optimization in an Integrated Sensing and Communication (ISAC) system, using a mix of OMA and NOMA to support many devices. A max-min problem was formulated to enhance the sensing beampattern with joint power allocation, beamforming, and RIS phase shift. The proposed low-complexity AO algorithm improved the beampattern gain and demonstrated trade-offs between sensing and communication. The authors in [90] examined a STAR-RIS-aided NOMA-ISAC system, focusing on minimizing matching error for the desired sensing beampattern through optimized beamforming and resource allocation. An AO algorithm, utilizing SCA and semidefinite relaxation, was applied to solve the complex problem iteratively until convergence. Simulation results confirmed the effectiveness of the approach. Another paper [91], the authors investigated a RIS-MIMO-NOMA system focusing on power consumption, proposing an inter-group NOMA scheme to reduce total transmit power, optimizing beamforming and phase shifts, and demonstrating through simulations that inter-group NOMA outperformed traditional clustering-based NOMA in power efficiency.

*Summary and Lessons Learned:* **Summary:** The RIS-enabled NOMA-based resource allocation frameworks presented in Table IV demonstrate a range of optimization techniques that significantly enhance system performance. These studies focus on both perfect and statistical CSI scenarios, integrating RIS configurations to optimize power allocation, beamforming, and phase shifts.

In uplink scenarios with perfect CSI, methods such as Trust Region-Based SCA and Penalty-Based SDR are used for STAR-RIS configuration and transmit power optimization [79]. Similarly, SCA, BCD, and Dinkelbach methods are applied to optimize power and beamforming vectors [87], while multiple RIS configurations leverage MMSE filters and iterative methods to refine beamforming and phase shifts [91]. For statistical CSI, energy consumption in NOMA-based MEC systems is optimized by managing transmission and reflection time [85], and sequential rank-one constraint relaxation is employed for joint resource allocation [84].

In downlink scenarios, optimization strategies for perfect CSI include power and discrete amplitude optimization [80], resource allocation using matching theory and AO [81], and a two-layer iterative algorithm for UE pairing and beamforming [88]. ISAC systems with perfect CSI optimize power and RIS phase shifts with AO and SCA methods [89], while STAR-RIS setups use semidefinite relaxation techniques [90]. For statistical CSI, algorithms are applied to optimize power allocation and RIS coefficients [82], and downlink URLLC systems manage beamforming and power with AO [83]. Additionally, multiple RIS configurations in downlink NOMA systems utilize AO and bisection methods for optimizing power and phase shifts [86].

However, a critical gap in the existing literature is the limited exploration of imperfect CSI scenarios. Most studies focus on perfect or known CSI, which may not reflect real-world conditions where channel state information is often imperfect. Additionally, the scalability and computational complexity of multi-RIS configurations present challenges that have yet to be fully addressed. Further research in these areas is essential to develop more robust and adaptable RIS-assisted NOMA frameworks capable of handling diverse and dynamic network environments effectively.

**Lessons Learned:** In the deployment of RIS-enabled NOMA systems, various factors such as deployment strategies, system configurations, and hardware elements play crucial roles in determining performance outcomes. The following insights summarize key lessons learned from integrating RIS with NOMA systems:

· Enhancing NOMA Performance with STAR-RIS: Utilizing STAR-RIS in NOMA systems can significantly improve performance by creating distinct channel conditions for different users. This differentiation allows the system to better exploit beamforming and multiplexing gains, resulting in enhanced signal strength and overall efficiency [81].

TABLE IV: Summary of NOMA-based Resource Allocation Frameworks

| Ref | Year | System Model | RIS Details | | | | | | | CSI | Design and Optimization | |
|---|---|---|---|---|---|---|---|---|---|---|---|---|
| | | | No's | Types | | | Modes | | | | Designed Variable | Optimization Methodology |
| | | | | A | P | H | R | T | STAR | | | |
| [79] | 2021 | UL, NOMA, BS (SA), UEs (SA) | Single | X | ✓ | X | X | X | ✓ | Perfect | The transmit power and configuration scheme at the STAR-RIS | Trust Region-Based SCA, and Penalty-Based SDR |
| [80] | 2022 | DL, NOMA, BS (SA), UEs (SA) | Single | X | ✓ | X | X | X | ✓ | Perfect | Power and discrete amplitude | A joint power and discrete amplitude allocation scheme |
| [81] | 2022 | DL, OMA and NOMA, AP (SA), UEs (SA) | Single | X | ✓ | X | X | X | ✓ | Perfect | Channel, power, and beamforming | Matching theory & AO |
| [82] | 2023 | DL, NOMA, AP (SA), UEs (SA) | Single | X | ✓ | X | X | X | ✓ | Statistical | Power allocation of UEs and amplitude coefficients of RIS elements | Three algorithms |
| [83] | 2023 | DL and UL, NOMA, URLLC, BS (MA), UEs (SA) | Single | X | ✓ | X | X | X | ✓ | Known | Beamforming and power | AO |
| [85] | 2023 | UL, NOMA, MEC, BS (MA), UEs (SA) | Single | X | ✓ | ✓ | X | X | ✓ | Known | Transmission and reflection time and coefficients of the STAR-RIS | An energy consumption optimization algorithm |
| [84] | 2023 | UL and DL, NOMA, BS (MA), UEs (SA) | Single | X | ✓ | X | ✓ | ✓ | X | Known | Sub-channel, decoding order, power, beamforming | AO, a penalty-based method, majorization minimization (MM) and semi-definite relaxation approaches, and a sequential rank-one constraint relaxation approach (SROCR) |
| [86] | 2023 | DL, NOMA, BS (MA), backscatter device (BD), UEs (SA) | Multiple | X | X | X | ✓ | ✓ | X | Known | Powers and phase shifts | AO & a bisection method |
| [87] | 2024 | UL, NOMA, BS (MA), Devices (SA) | Single | ✓ | ✓ | X | X | X | X | Perfect | Power and beamforming vectors | SCA, BCD and Dinkelbach methods |
| [88] | 2024 | DL, NOMA, BS (SA), UEs (SA) | Single | X | X | X | ✓ | ✓ | X | Known | UEs pairing, decoding order, passive beamforming, power and time | A novel two-layer iterative algorithm |
| [89] | 2024 | DL, NOMA, ISAC, BS (MA), UEs (SA), Radar Targets | Single | X | X | ✓ | X | X | X | Perfect | Power, active beamforming and RIS phase shift | A low complexity AO algorithm, penalty and SCA methods |
| [90] | 2024 | DL, NOMA, ISAC, BS (MA), UEs (SA), Radar Targets | Single | X | ✓ | X | X | X | ✓ | Knwon | The BS active beamforming, STAR-RIS passive beamforming, power and time | An AO algorithm basd on SCA and semidefinite relaxation techniques |
| [91] | 2024 | UL, NOMA, MIMO, BS (MA), UEs (SA) | Mutiple | X | X | ✓ | X | X | X | Perfect | Beamforming and phase shifts | The MMSE filters, iterative method and sequential rotation (SR) scheme |

· Critical Role of Channel Assignment: Effective channel assignment is essential in STAR-RIS-aided NOMA systems as it influences how users are paired and how signals are processed. By carefully assigning channels, the system can maximize its degrees of freedom, leading to better beamforming, improved multiplexing, and ultimately higher network efficiency [81].

· Decoding Order's Impact on NOMA: The order in which NOMA signals are decoded plays a significant role in system performance. Well-structured decoding strategies can provide substantial performance improvements compared to random or less methodical approaches. Although exhaustive search algorithms offer optimal results, more practical, structured decoding orders are necessary for maintaining efficiency in larger networks [81].

Increasing Antennas at the BS: Adding more antennas to the BS enhances its ability to perform beamforming, which boosts the system's weighted sum-rate. NOMA systems, which are designed to serve multiple users simultaneously, benefit greatly from this increase. Moreover, combining passive and active beamforming optimizations leads to even greater performance improvements compared to conventional methods [84].

· Managing Transmit Power and System Efficiency: While increasing transmit power at the BS can improve downlink performance and the weighted sum-rate, it also raises the risk of self-interference, particularly in full-duplex systems. This self-interference can degrade uplink performance, highlighting the need for careful power management to balance the benefits and drawbacks effectively [84].

· Balancing RIS Elements and Transmit Power: As the number of reconfigurable elements in an RIS increases, the system can achieve better channel conditions with lower transmit power. This improvement is due to the additional flexibility provided by more elements, allowing the system to optimize performance while conserving energy [91].

· Advantages of NOMA Over OMA: NOMA systems offer significant benefits over OMA by enabling multiple users to share the same time-frequency resources, which increases spectral efficiency and overall system performance. Systems that strategically optimize phase adjustments perform better than those using random phase settings, underscoring the importance of deliberate phase control in achieving optimal results [91].

### D. Resource Allocation for THz Communication Networks with RIS

This subsection delves into resource allocation strategies in RIS-aided THz networks, emphasizing the optimization of key design variables such as beamforming, power allocation, phase shifts, and user scheduling to boost system performance, as shown in Figure. 7. The reviewed studies tackle challenges like maximizing system throughput, improving EE, and enhancing secure communications in the THz band. Various approaches, including single and multiple RIS configurations, are employed, leveraging techniques such as BCD, coalition game theory, and iterative algorithms to optimize these variables. The integration of RIS with massive MIMO technologies is highlighted for its potential to enhance system capacity and EE, despite the challenges of imperfect CSI and the highly directional nature of THz signals. However, managing multiple RISs, mitigating beam split effects, and ensuring robust performance against security threats, such as eavesdropping, remain significant challenges. These findings underscore the importance of advanced optimization techniques in realizing the full potential of RIS-aided THz networks, especially in dynamic and high-frequency communication environments.

This paper [92] addressed resource allocation in a cellular network with RIS integration, presenting a model covering multiple BSs and user scenarios. The goal was to maximize the system's total rate by dividing the challenge into user resource allocation and RIS phase shift optimization. Using BCD and coalition game and discrete phase search algorithms, the authors demonstrated through simulations that their approach significantly improved the system's rate while keeping complexity low. In another paper [93], the authors focused on optimizing resource allocation in the THz band to enhance EE. The authors established an RIS-assisted THz-MIMO downlink wireless communication framework, addressing EE improvement through phase-shift matrix refinement and power distribution optimization. Their decentralized EE optimization algorithm simplified the complex problem into a tractable convex optimization task, and simulations showed that it converged quickly and reached optimal EE. The work validated the practicality and effectiveness of integrating RIS and massive MIMO technologies in THz communication networks.

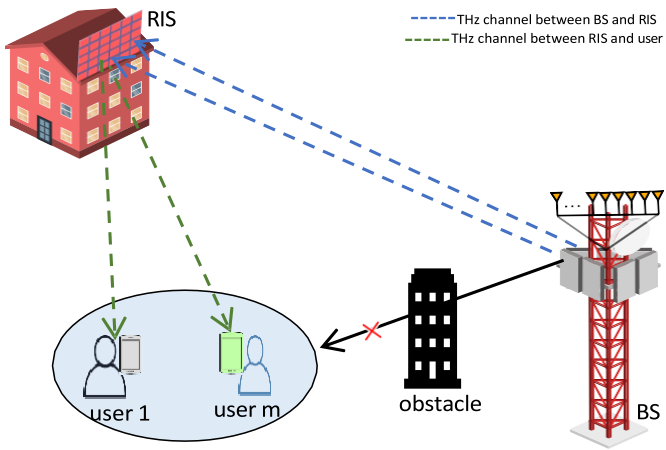

Fig. 7: Illustrating RIS-assisted THz-MIMO downlink wireless networks.

Contrasting the single RIS schemes in [92], [93], this paper [94] focused on multiple RIS for enhancing secure communications in THz systems, amidst potential eavesdropping risks. The study analyzed blocking effects on transmissions due THz bands' blockage-prone nature and derived the worst-case secrecy rate under imperfect channel information. It suggested methods to secure BS-RIS and RIS-user links: employing zero-forcing hybrid beamforming for the former and using an iterative algorithm for the latter. The research also addressed multi-Eves scenarios with a robust transmission strategy. Simulations indicated that their approach significantly improved secrecy performance, overcoming the blockage issues in THz communication.

This paper [95] addressed the beam split problem in THz RIS-aided 6G communications and investigated the effects of RIS size, shape, and placement on beam splitting. A hybrid beamforming architecture was implemented at the BS, using distributed RISs to mitigate array gain loss. The study aimed to maximize the sum rate by optimizing beamforming, time delays, and reflection coefficients. An iterative algorithm, incorporating techniques like LDR and MCQT, was used to optimize these parameters. Simulations showed that this approach effectively reduced beam split effects and enhanced system capacity. This work [96] investigated a MU-MIMO THz communication system in an indoor setting where obstacles block direct links. Using a nearby RIS with adjustable phase shifts, the study formulated a problem to maximize achievable rates. The accelerated proximal gradient method was applied, effectively handling the non-convex constraints through normalization. Results demonstrated that this approach improved range by 30% to 120% for achieving 100 Gbps with 81 RIS elements, even with phase shift quantization and imperfect channel knowledge. In another paper [97], the authors proposed a framework for THz wireless Virtual Reality (VR) systems using multi-RIS to enhance QoE. Formulated an optimization problem to maximize QoE by jointly optimizing RIS beamforming, VR user power allocation, and virtual object rendering capacity. They divided the optimization into two stages: Bit Error Rate (BER) and maximizing rate, followed by maximizing rendering capacity. Simulations confirmed the framework's superior QoE performance compared baselines.

*Summary and Lessons Learned:* **Summary:** The studies highlighted in Table V show considerable progress in optimizing resource allocation for THz communication networks through RIS integration. Research efforts such as [92] and [93] focused on improving system performance and EE in single RIS setups, utilizing algorithms like BCD and coalition games for resource management. Additionally, [94] explored the use of multiple RIS configurations to enhance security in THz systems, employing hybrid beamforming techniques. While these studies demonstrate effective strategies, they often assume ideal conditions, such as perfect or accurate CSI, which may not reflect real-world scenarios.

Despite these advancements, there are critical gaps that need addressing. Many studies overlook the challenges associated with imperfect CSI, a common issue in practical deployments that can significantly impact performance. Furthermore, the complexity and scalability of managing multiple RIS configurations, particularly in dynamic environments typical of THz networks, are not fully explored. The integration of RIS into applications like VR, as investigated in [97], shows potential but also highlights the need for more sophisticated optimization methods that can meet the demanding requirements of high-speed, low-latency applications. Bridging these gaps is crucial for the successful deployment of RIS-assisted THz networks in real-world conditions. Overall, the complexity of managing multiple RISs, mitigating beam split effects, and ensuring robust performance under security threats like eavesdropping remains a critical area for further research.

**Lessons Learned:** In THz networks, lessons include managing impact on interference in mmWave systems, securing communications against eavesdropping and blockages in THz bands, and employing MIMO technology to enhance signal reach and robustness despite severe attenuation and diffraction. Following are the insights from implementing RIS for THz networks:

· Impact of Beamwidth: In mmWave networks, increasing the angle of the user-power beamwidth generally leads to a decrease system sum rate. This is due to the wider coverage area causing more interference to other users. Most algorithms show a decline in performance as this angle increases, with the interference effect eventually nearing saturation [92].

· Securing THz Communication: Secure communication in THz bands is highly susceptible to blockage and eavesdropping, with power loss from eavesdropper blocking being a critical factor. Key lessons learned include the significant impact of blockage on secrecy rates and the importance of designing secure transmission strategies that account for potential eavesdroppers and their blocking effects. Effective countermeasures must consider both the direct interception of signals and the additional blockages introduced by eavesdroppers [94], [94].

TABLE V: Summary of THz-based Resource Allocation Schemes

| Ref | Year | System Model | RIS Details | | | | | | | CSI | Design and Optimization | |
|---|---|---|---|---|---|---|---|---|---|---|---|---|
| | | | No's | Types | | | | Modes | | | Designed Variable | Optimization Methodology |
| | | | | A | P | H | R | T | STAR | | | |
| [92] | 2023 | UL, THz, HetNets, BSs (MA), UEs (SA) | Single | ✗ | ✓ | ✗ | ✓ | ✗ | ✗ | Known | UEs resource allocation and the phase shift | BCD and coalition game and discrete phase search algorithms |
| [93] | 2021 | DL, THz, MIMO, BS (MA), UEs (SA) | Single | ✗ | ✓ | ✗ | ✓ | ✗ | ✗ | Known | RIS phase-shift and power | Distributed EE optimization algorithm |
| [94] | 2022 | DL, THz, MIMO, BS (MA), UEs (SA) | Multiple | ✗ | ✓ | ✗ | ✗ | ✓ | ✗ | Imperfect | Beamforming and phase shifts | Zero-forcing hybrid beamforming and an iterative algorithm |
| [95] | 2023 | DL, THz, BS (MA), UEs (SA) | Single | ✗ | ✓ | ✗ | ✓ | ✗ | ✗ | Known | Hybrid analog/digital beamforming, time delays at the BS and reflection coefficients at the RISs | MMSE and alternating direction method of multipliers (ADMM) |
| [96] | 2023 | DL, THz, MU-MIMO, BS (MA), UEs (SA) | Single | ✗ | ✓ | ✗ | ✓ | ✗ | ✗ | Imperfect | Phase shifts | The accelerated proximal gradient (APG) method |
| [97] | 2024 | DL, THz, MIMO, MEC, AP (MA), VR UEs (SA) | Multiple | ✓ | ✗ | ✗ | ✗ | ✓ | ✗ | Known | The passive beamforming at RIS, the transmit power allocation among VR users, and the rendering capacity allocation among virtual objects | Objective function conversion and AO methods |

- Enhancing THz Signal Transmission with MIMO: THz signals typically suffer from severe attenuation and poor diffraction, resulting in limited coverage and short transmission distances. To address these issues, MIMO technology can be applied to generate high-gain directional beams. By using large antenna arrays, MIMO technology significantly enhances signal strength and directionality, allowing the signal to cover larger areas and travel longer distances. High-gain directional beams not only reduce signal attenuation but also overcome poor diffraction, thereby improving overall communication performance [95].

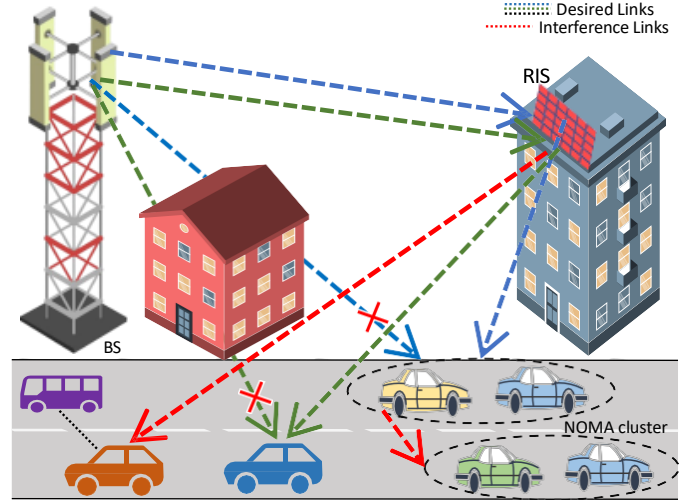

Fig. 8: Depiction of RIS-assisted vehicular network.

### E. RIS-Driven Resource Allocation in Vehicular Communication Networks

In this subsection, our attention is directed towards the application of RIS for resource allocation optimization in communication networks designed for vehicles. In this specific context, RIS assumes a pivotal role in elevating the effectiveness and performance of wireless communication systems specifically designed to cater to the unique requirements of vehicular applications. This particular area of research into the integration of RIS into vehicular networks, with the primary objectives of enhancing connectivity, mitigating interference, and overall, enhancing the quality of communication services for mobile vehicles. Recent advancements are detailed in Table VI. The RIS MISO NOMA-enabled vehicular system is demonstrated in Figure. 8.

This paper [98] examined resource allocation in RIS-assisted VC networks, focusing on the use of large-scale fading channel information. The authors optimized the network for varying QoS demands, aiming to maximize Vehicle-to-Infrastructure (V2I) link capacity while ensuring baseline SINR for Vehicle-to-Vehicle (V2V) connections. They tackled the mixed integer, non-convex optimization problem in two stages using an AO algorithm. Simulations demonstrated that this approach significantly improved V2I link capacity, highlighting RIS's critical role in meeting stringent QoS requirements in VC networks. Similarly, in [99], the authors aimed to improve VC by using a hybrid OMA/NOMA scheme with an RIS in dynamic settings. It focused on maximizing the SSR by optimizing transmit power, RIS phase shifts, and vehicle beamforming, considering various QoS levels. The optimization was split into three subproblems and tackled with an iterative SCA-based algorithm. Simulations revealed that this method significantly enhanced the SSR, especially in cell-edge zones, and was effective at various vehicle speeds.

Unlike the works in [98], [99] that considered single RIS and perfect CSI, the authors in this paper [100] assumed multiple RISs and imperfect CSI to address challenges in RIS-aided VC. They introduced an active RIS to boost signals and improve connectivity for vehicles. It focused on optimizing transmit power and RIS coefficients, using two algorithms AO and Constrained Stochastic Successive Convex Approximation (CSSCA) to streamline the process. Simulations demonstrated that active RIS significantly enhanced communication performance. In another paper [101], the authors introduced a UAV-enhanced RIS-assisted V2X communication architecture (UR-V2X) for urban IoT traffic, featuring a specialized MAC protocol (UR-V2X-MAC) to optimize resource allocation and scheduling. The architecture utilizes UAVs as access points and RISs as passive relays to enhance communication capacity and reduce latency. Using a distributed optimization algorithm for power and RIS phase shift adjustments, UR-V2X-MAC demonstrated reduced delays and increased system capacity compared to traditional V2X protocols and non-RIS setups.

*Summary and Lessons Learned:* **Summary:** These advancements, summarized in Table VI, demonstrate the critical role of RIS in optimizing resource allocation and enhancing performance in VC networks. Resource allocation schemes for VC primarily utilize single RIS setups with perfect CSI to optimize power, reflection coefficients, and beamforming. For instance, [98] explored downlink V2I, V2V, and D2D communication, optimizing power and RIS coefficients using the AO method. Similarly, [99] examined a hybrid NOMA/OMA system, focusing on transmit power and RIS phase shift optimization using the SCA algorithm. In [100], multiple RISs for downlink systems are considered to optimize BS transmit power through AO and CSSCA algorithms. Additionally, active RIS was considered in [99] and [100], focusing on enhancing signal strength and connectivity in vehicular networks. These approaches highlight the diverse methodologies used to enhance vehicular communication performance, catering to different CSI conditions and system configurations.

**Lessons Learned:** Strategic RIS deployment in vehicular networks compensates for channel gain loss due to vehicle mobility, enhancing signal strength. Balancing RIS elements optimizes performance without adding complexity. Selecting NOMA and OMA access technologies addresses different service needs effectively. Integrating UAVs into V2X communications improves reliability and coverage, indicating a beneficial area for future research. Here are the insights:

- Effective RIS Deployment: Deploying RIS effectively in vehicular networks is crucial to mitigate the challenges posed by vehicle mobility, such as channel gain loss. By strategically placing RIS, it's possible to significantly enhance signal strength, ensuring stable communication even for vehicles that are farther from the BS. This approach helps maintain robust connectivity across the network despite the high mobility of vehicles [98].
- Balancing Performance and Complexity with RIS Elements: While adding more RIS elements can boost system performance, it also introduces additional hardware complexity and increases pilot overhead. Therefore, it's important to strike a balance between the performance benefits and the practical challenges. In some scenarios, optimizing the power budget may be a more effective strategy than simply increasing the number of RIS elements. This careful balancing act is essential for designing efficient and practical systems [99].
- Optimizing Access Technology Choices: The integration of NOMA and OMA access technologies within vehicular networks offers a flexible solution to meet diverse service demands. OMA is particularly suitable for scenarios requiring high reliability, such as emergency vehicle communications, through careful bandwidth allocation. Conversely, NOMA enhances spectral efficiency by allowing multiple users with different QoS requirements to share the same spectral resources, which is particularly beneficial in crowded network environments [99].
- Leveraging UAVs for Enhanced V2X Communications: Integrating UAVs into V2X communication systems can greatly improve both coverage and reliability. UAVs, with their ability to establish direct LoS connections, help maintain consistent communication links, even in challenging environments. Deploying multiple UAVs can further extend coverage, ensuring uninterrupted service across broader areas. This presents a promising avenue for future research to enhance the capabilities of vehicular networks [101].

### F. RIS-Integrated Resource Allocation for UAV Communication

This subsection explores resource allocation strategies in RIS-integrated UAV communication systems, emphasizing the optimization of key design variables such as UAV trajectories, power allocation, RIS phase shifts, beamforming, and user scheduling. The studies reviewed employ diverse approaches, including single and multiple RIS setups, to address challenges in cognitive radio systems, NOMA networks, and MEC-enabled UAV systems. Methods like iterative optimization, SCA, BCD, and Dinkelbach's method are utilized to enhance system performance, leading to benefits such as increased throughput, reduced power consumption, and improved secure communication. Despite these advancements, challenges remain in scaling these solutions and effectively managing imperfect CSI in dynamic and multi-user environments. Future research should focus on developing more robust strategies that can better adapt to real-world complexities, ensuring the practical implementation and scalability of these technologies.

The paper [102] introduced a joint optimization strategy to increase throughput in UAV-enabled cognitive radio systems. Using RIS technology, the study tackled interference issues from spectrum sharing and developed an iterative optimization algorithm for suboptimal solutions. Numerical results demonstrated notable throughput enhancements through this method. Unlike single RIS-based work in [102], the authors in this study [103] considered multiple RIS with statistical CSI and investigated UAV-assisted NOMA network to minimize power consumption and meet data rate requirements. It addressed optimization of UAV positions, RIS coefficients, and power management, using several optimization techniques. Simulation results showed reduced power consumption and highlighted RIS benefits in multi-UAV networks.

This paper [104] focused on improving EE in UAV-enabled MEC systems with RIS support, optimizing bit allocation, transmit power, phase shift, and UAV trajectory. An iterative algorithm using Dinkelbach's method and BCD enhanced EE and met latency demands. Simulations showed the impact of mission period, task input, and channel state on EE. Similarly, the authors in [105] introduced a secure communication strategy for UAV-MEC systems utilizing RIS, aimed at boosting security with RIS-based jamming techniques and enhancing computing capacity through optimal resource and trajectory management. Simulation results confirmed the strategy's superiority in enhancing secure computing performance over existing benchmark approaches.

In another paper [106], the authors examined a RIS-enhanced MIMO system with a UAV, focusing on maximizing the average WSR by optimizing user scheduling, UAV trajectory, and beamforming strategies. The complex problem was divided into subproblems, tackled through conic relaxation, alternating direction method of multipliers, and SCA. Numerical results validated the effectiveness of the approach in enhancing rates within backhaul capacity limits.

TABLE VI: Summary of Resource Allocation Schemes for Vehicular Communication Networks

| Ref | Year | System Model | RIS Details | | | | | | | CSI | Design and Optimization | |
|-----|------|--------------|-------------|---|---|---|---|---|---|-----|-------------------------|---|
| | | | No's | A | P | H | R | T | STAR | | Designed Variable | Optimization Methodology |
| [98] | 2020 | DL, V2I, V2V, D2D, BS (SA), Vehicle UEs (SA) | Single | X | ✓ | X | ✓ | X | X | Perfect | Power, RIS reflection coefficients and spectrum | AO |
| [99] | 2022 | DL, V2I, V2V, NOMA & OMA, BS (MA), Vehicles (SA) | Single | ✓ | X | X | ✓ | X | X | Perfect | Transmit power, RIS's phase shift, and beamforming | SCA |
| [100] | 2022 | DL, BS (MA), UEs (SA) | Multiple | ✓ | X | X | X | X | ✓ | Imperfect | The BS's transmit power and RIS coefficients | AO & CSSCA algorithm |
| [101] | 2024 | DL, V2X, intelligent connected vehicles (ICVs), UAV | Multiple | X | ✓ | X | X | ✓ | X | Known | The transmit power and the RIS phase shift | The distributed AO method |

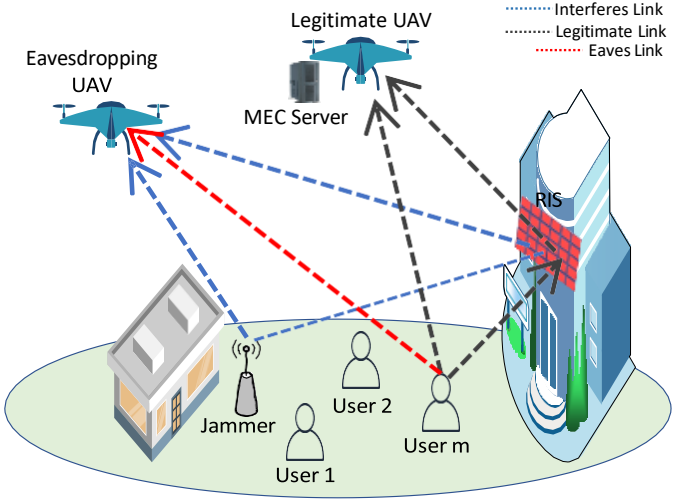

Fig. 9: RIS-assisted UAV-MEC secure computing system

*Summary and Lessons Learned:* **Summary:** Table VII highlights recent advancements in the application of RIS for optimizing resource allocation in UAV communication systems. Most studies, such as [102], focus on single RIS setups. In this particular study, a downlink UAV system involving primary and secondary UEs is analyzed, where statistical CSI is used to optimize UAV trajectory, RIS passive beamforming, and power allocation. Similarly, [104] examines an uplink UAV system with MEC and NOMA IoT devices, employing imperfect CSI to refine bit allocation, transmit power, phase shifts, and UAV trajectory. The use of multiple RIS setups is less common but is explored in [103], which looks at a downlink UAV NOMA system. This research uses statistical CSI to optimize the positions of UAVs, RIS reflection coefficients, transmit power, and active beamforming vectors. Another study, [105], investigates an uplink MEC system with a mix of Eve UAVs, a legitimate UAV, a jammer, and UEs, utilizing known CSI to optimize variables like frame length, transmit power, local computational data allocation, UAV trajectory, and RIS phase shifts. Additionally, [106] explores a downlink UAV MIMO system with BS and UEs, applying known CSI to optimize user scheduling, UAV trajectory, and both active and passive beamforming. These studies demonstrate the effectiveness of RIS in enhancing UAV communication systems by optimizing key parameters such as trajectory, power allocation, and beamforming under various CSI conditions. While single RIS setups are predominantly studied, there is limited explo-ration of multiple RIS configurations, indicating a potential area for future research.

**Lessons Learned:**

· Optimizing RIS Placement: Placing RIS units closer to UAVs enhances performance by improving passive beamforming through better phase-shift control. This strategic positioning leads to reduced power consumption and more effective resource management in UAV communication systems [103].

· Altitude Challenges and Power Use: Higher UAV flight altitudes can lead to greater channel attenuation and weaker signal reception, which increases power consumption in resource allocation. To address this, UAVs may require more power to maintain effective beamforming and ensure strong system performance at these higher altitudes [103].

· Enhancing Security with RIS: As data offloading to UAVs increases, so does the risk of data interception. By equipping RIS with more reflective elements, the system's ability to control the signal environment is enhanced, which strengthens security against potential eavesdropping [105].

### G. Resource Allocation in RIS-Aided Edge Networks

This subsection explores the optimization of resource allocation in edge networks augmented by RIS. The emphasis is on how RIS can enhance the distribution of resources such as bandwidth and computational capacity, particularly in scenarios where edge computing is pivotal. The discussion underscores the benefits of RIS in these networks, including increased efficiency, reduced latency, and overall performance improvements. These enhancements are depicted in Figure. 10, which illustrates the influence of RIS on resource management within edge networks. The subsection concludes with a summary of key insights and lessons learned, accompanied by a comparative summary table that offers a comprehensive view of the advanced schemes and their effectiveness in optimizing resource allocation in RIS-aided edge networks.

This research [107] investigated integrating RIS into Wireless Powered Mobile Edge Computing (WP-MEC) systems to enhance computational capabilities and EE, especially under poor transmission conditions. It proposed optimizing power allocation, computing frequencies, and RIS reflection coefficients in a multi-user Orthogonal Frequency Division Multiplexing (OFDM) scenario. Using AO and SCA, the approach significantly reduced energy consumption—by 80% compared to traditional solutions—proving effective in simulations. Another work [108] explored how RIS could enhance MEC by reshaping wireless channels. It focused on optimizing RIS phase shifts and resource allocation in a multi-hop MEC network. Using spectral graph theory, the researchers developed an iterative algorithm, demonstrating through simulations that RIS significantly enhanced network throughput.

TABLE VII: Summary of Resource Allocation Schemes for UAV

| Ref | Year | System Model | RIS Details | | | | | | | CSI | Design and Optimization | |
|---|---|---|---|---|---|---|---|---|---|---|---|---|
| | | | No's | Types | | | Modes | | | | Designed Variable | Optimization Methodology |
| | | | | A | P | H | R | T | STAR | | | |
| [102] | 2023 | DL, UAV, Primary and Secondary UEs (MA), Primary and Secondary BS (MA) | Single | ✗ | ✓ | ✗ | ✓ | ✗ | ✗ | Known | UAV's trajectory, RIS's passive beamforming and UAV's power allocation | An alternating iterative optimization algorithm |
| [103] | 2023 | DL, UAV (MA), NOMA, UEs | Multiple | ✗ | ✓ | ✗ | ✓ | ✗ | ✗ | Known | The position of UAVs, RIS reflection coefficients, transmit power, active beamforming vectors and decoding order, and thus is quite hard to solve optimally | Gaussian randomization & effective convex optimization techniques |
| [104] | 2023 | UL, UAV, MEC, NOMA, IoT devices | Single | ✗ | ✓ | ✗ | ✓ | ✗ | ✗ | Imperfect | Bit allocation, transmit power, phase shift, and UAV trajectory | Dinkelbach's method and BCD technique |
| [105] | 2023 | UL, MEC, Eve UAVs (MA), Legitimate UAV (MA), Jammer (MA), UEs (MA) | Single | ✓ | ✗ | ✗ | ✓ | ✗ | ✗ | Known | The frame length, transmit power, local computational data allocation, UAV trajectory, and RIS phase shift | SCA & BCD |
| [106] | 2024 | DL, UAV, MIMO, BS (MA), UEs (SA) | Single | ✗ | ✓ | ✗ | ✓ | ✗ | ✗ | Known | The user scheduling, UAV's trajectory, active beamforming and passive beamforming | The conic relaxation, the alternating direction method and SCA |

In contrast to the above works [107], [108] focusing on a single RIS scenario, the authors in [109] considered multiple RIS and introduced a dynamic algorithm to improve EE and reduce latency in MEC frameworks. It optimized radio and computational resources, and RIS reflectivity parameters to meet specific performance goals. Operating on a per-slot basis, the algorithm effectively handled varying channels and unpredictable task arrivals. Numerical results confirmed its effectiveness, showing notable improvements in EE and latency. Similarly, this paper [110] introduced an algorithm to enhance EE and reduce latency in dynamic MEC within B5G networks, using RISs. Leveraging stochastic optimization, the authors developed a dynamic learning algorithm to optimize resources and RIS parameters for specified end-to-end delay targets. This strategy provided dynamic control, even amidst uncertain radio channels and task arrivals, crucially improving dynamic MEC performance with RIS inclusion and optimization, as validated by numerical results. In another study [111], the authors investigated intelligent B5G vehicular networks, highlighting MEC's role in low-latency applications. It analyzed RIS-assisted MEC-served vehicular networks, focusing on task scheduling optimization. A dynamic algorithm was proposed and validated through simulations, showing improved task offloading, computing, and completion rates compared to benchmarks. This work addressed mmWave signal attenuation challenges and enhanced task scheduling efficiency in vehicular networks.

This study [112] investigated how MEC and RIS enhance processing in power grids. It aimed to reduce energy consumption by optimizing key parameters such as transmission power and computing resources. Using various mathematical transformations, the research confirmed that RIS integration with MEC effectively lowers energy usage in power networks. In another paper [113], the authors introduced RISs in multi-server MEC systems to enhance performance. By jointly optimizing user association, beamforming, and resource allocation, the authors aimed to maximize the task completion rate. The proposed algorithms outperformed benchmark schemes, showcasing the potential benefits of integrating RISs into MEC systems. Similarly, authors in [114] proposed a joint optimization approach for resource allocation in an RIS-assisted MEC network. Their method considers the tradeoff between offloading delay and energy consumption. By decou-pling the original problem and employing SCA methods, they effectively solved the optimization problem. Simulation results illustrated the superiority of their method in terms of cost savings compared to benchmark methods.

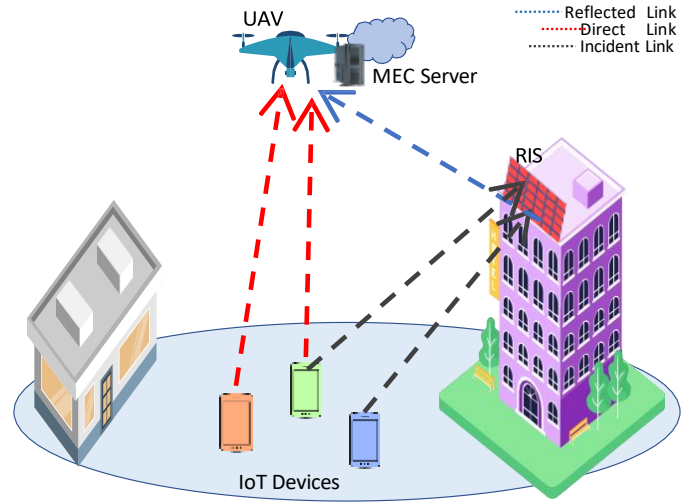

Fig. 10: Depiction of RIS-assisted UAV-enabled MEC system.

Different from the works [107]–[114], this paper [115] introduced STAR-RIS to enhance energy transfer and task offloading in wireless-powered MEC systems. Three operating protocols—Energy Splitting (ES), Mode Switching (MS), and Time Splitting (TS)—were studied. An iterative algorithm was proposed for the ES protocol, extending to solve MS and TS problems. Simulation results showed STAR-RIS outperforming traditional RIS, with the TS protocol achieving the highest computation rate.

This paper [116] tackled network-wide latency optimization in MEC systems using multi-RIS-assisted offloading. The problem was split into path selection, to balance beamforming gain and reflection loss, and joint RIS phase and offloading optimization using SDR for efficiency. Simulations revealed significant latency reductions—24% over RIS-agnostic systems and 15.34% compared to single RIS configurations.

Distinct from all the aforementioned OMA-based works in the category [107]–[116], this article [117] considered NOMA-based cooperative MEC network with SWIPT. They focused on minimizing task delay and energy consumption. The network design included joint optimization of the RIS's reflection beamforming vector, power splitting ratio, relay power allocation, and task offloading parameters. An iterative optimization algorithm utilizing SCA and a penalty method was developed to address these challenges, showing through simulations that the proposed approach outperformed conventional orthogonal MEC networks in reducing delay and energy use. Another paper [118] addressed power minimization in an uplink multi-RIS assisted multi-cell NOMA network, introducing a NOMA scheme with Inter-Group Interference Cancellation (IGIC) to reduce user interference. The approach involved optimizing equalizers, transmit power, and RIS phase shifts. Simulations showed that this scheme, using IGIC and multi-reflection RIS, effectively lowered total transmit power compared to other strategies.

*Summary and Lessons Learned:* **Summary:** The resource allocation schemes for RIS-assisted edge networks are explored across various system models, encompassing both single and multiple RIS configurations, with a focus on utilizing known and perfect CSI, as shown in Table VIII. Studies on single RIS setups, such as those by [107], [108], and [112], emphasize optimizing signals, frequencies, power, and RIS phase shifts. In addition, works like [114], [115], and [117] focus on optimizing time, transmit power, CPU frequencies, and signal detection vectors in single RIS configurations with perfect CSI. For multiple RIS configurations, known CSI is investigated in studies like [116] and [118], while perfect CSI is leveraged in [109], [111], and [113] to optimize radio resources, computation resources, and RIS reflectivity parameters. These studies highlight the importance of improving processor resource allocation, offloading policies, and beamforming at both BSs and RISs. Active RIS also plays a crucial role in several schemes, particularly in [107], [113], [114], and [116], where the emphasis is on optimizing transmission power, computing resources, and RIS phase shifts.

Despite these advancements, significant gaps remain in the exploration of imperfect CSI within these frameworks. The current focus on known and perfect CSI overlooks the challenges posed by imperfect CSI, which is common in real-world scenarios. This gap is particularly critical as the accuracy of CSI directly impacts the efficiency of resource allocation strategies. Moreover, the scalability and complexity of managing multi-RIS deployments pose additional challenges, as optimizing power allocation, phase shifts, and beamforming becomes increasingly difficult with more RIS elements. Furthermore, the integration of RIS into heterogeneous networks remains underexplored, requiring strategies for seamless interoperability across different network technologies and devices. The dynamic nature of edge networks also calls for real-time adaptation mechanisms that can respond to changing conditions, such as user mobility and varying traffic loads. Additionally, while RIS can enhance communication efficiency, the overall energy impact on edge devices, especially those with limited power budgets, needs further examination, as the computational demands introduced by RIS optimization may offset energy savings. Addressing these gaps through targeted research is crucial for the effective deployment of RIS in edge networks. Developing robust algorithms that account for imperfect CSI, scalable optimization frameworks for multi-RIS deployments, and strategies for seamless integration and real-time adaptation will significantly enhance the robustness, efficiency, and adaptability of RIS-assisted edge networks in diverse and dynamic environments.

**Lessons Learned:**

- Striking a Balance: In wireless communication systems, achieving a balance between computational complexity, energy consumption, and latency is essential [111]. MEC addresses these challenges by offloading tasks from devices to edge servers, thereby improving processing efficiency and reducing energy usage. This method accelerates data processing and decreases dependence on distant cloud services, but it requires a robust network infrastructure to support real-time operations [114].

- Impact of Path Loss Exponent: The path loss exponent is a key factor that can significantly affect communication performance, particularly by increasing offloading delays and energy consumption. The benefits of RIS become more evident under challenging channel conditions, where they can help mitigate the negative effects of path loss [114].

- EE in Computation Offloading: The integration of RIS in MEC environments has been found to significantly reduce the energy consumption of IoT devices by enhancing wireless signal propagation and reducing transmission power needs [111]. However, it is critical to balance the energy savings from offloading tasks to edge servers with the energy required to operate the RIS. Detailed evaluations of RIS energy consumption across different operational states and the development of dynamic algorithms for RIS activity adjustment are essential [114]. Effective smart task distribution between IoT devices and edge servers also plays a key role in optimizing overall EE. Additionally, considering varying signal propagation characteristics and diverse energy profiles of IoT devices is vital for creating an energy-efficient RIS-assisted computational offloading framework in MEC environments [115], [116].

### H. AI-based RIS-Aided Resource Allocation Optimization

This subsection reviews AI-driven frameworks for optimizing resource allocation in RIS-aided UAV and MEC systems, spanning various network types such as SISO, MISO, MIMO, HetNets, NOMA, mmWave, THz, VC, and edge networks. The focus is on optimizing critical design variables, including beamforming vectors, power allocation, RIS phase shifts, and UAV trajectories, using advanced AI techniques like reinforcement learning (RL) and deep learning. Key studies have demonstrated the effectiveness of these AI methods in optimizing design variables as shown in Figs. 11 and 12. For example, in MISO systems, AI-driven optimization of beamforming and RIS phase shifts has led to improved user performance and reduced latency. Similarly, in THz and mmWave communications, optimizing hybrid beamforming, power allocation, and RIS configurations has resulted in enhanced spectral efficiency and lower latency. However, challenges remain, particularly in scaling these solutions across multiple RISs and ensuring adaptability in dynamic environments. Further research is needed to address issues such as imperfect CSI, computational complexity, and the deployment of AI in large-scale networks.

TABLE VIII: Summary of Resource Allocation Frameworks for Edge Networks

| Ref | Year | System Model | RIS Details | | | | | | | CSI | Design and Optimization | |
|-----|------|-------------|-------|---|---|---|---|---|------|-----|------------------------|--------------------------|
| | | | No's | A | P | H | R | T | STAR | | Designed Variable | Optimization Methodology |
| [107] | 2021 | UL, MEC, OFDM, Hybrid-AP (SA), Devices (SA) | Single | ✓ | ✗ | ✗ | ✓ | ✗ | ✗ | Known | Signals, frequencies, and the power | AO & SCA |
| [108] | 2021 | UL, MEC, MDs (SA) | Single | ✗ | ✓ | ✗ | ✓ | ✗ | ✗ | Known | RIS phase-shifts and the relays | A joint phase-shifts, power allocation, and bandwidth allocation optimization algorithm |
| [109] | 2021 | UL, Edge Server, AP (MA), Exde devices (MA) | Multiple | ✗ | ✓ | ✗ | ✓ | ✗ | ✗ | Perfect and Imperfect | radio resources (i.e., power, rates), computation resources (i.e., CPU cycles), and RIS reflectivity parameters (i.e., phase shifts) | A dynamic joint optimization algorithm |
| [110] | 2022 | UL and DL, Edge Server, AP (MA), Edge devices (MA) | Multiple | ✗ | ✓ | ✗ | ✓ | ✗ | ✗ | Static | Radio and computation resources | A novel algorithm for energy-efficient low-latency dynamic edge computing |
| [111] | 2022 | UL, MEC, V2I, RSU (MA), Vehicles UEs (MA) | Multiple | ✗ | ✓ | ✗ | ✓ | ✗ | ✗ | Perfect | Processor resource allocation and offloading policy | A dynamic task scheduling algorithm |
| [112] | 2022 | UL, MEC, BS (MA), MDs (SA) | Single | ✗ | ✓ | ✗ | ✓ | ✗ | ✗ | Known | The transmission power of MDs, the receive beamforming vector of BS, computing resource allocation, and the phase shift of RIS | Quadratic transformation and Lagrange dual transformation |
| [113] | 2023 | UL, MEC, BSs (MA), MDs (MA) | Multiple | ✓ | ✗ | ✗ | ✓ | ✗ | ✗ | Perfect | The MD-server association, the receive beamforming at BSs, the passive beamforming at RISs, and the computing resource allocation on edge servers | BCD, the penalty dual decomposition (PDD) method and a swap matching based algorithm |
| [114] | 2023 | UL, MEC, BS (MA), Mobile Terminals (SA) | Single | ✗ | ✓ | ✗ | ✓ | ✗ | ✗ | Perfect | The edge computing resource allocation, signal detecting vector, UL transmission power, and RIS phase shift coefficient | SCA |
| [115] | 2023 | UL and DL, MEC, AP (MA), UEs (MA) | Single | ✗ | ✓ | ✗ | ✗ | ✗ | ✓ | Perfect | Time, transmit power and CPU frequencies | SCA & iterative algorithm |
| [116] | 2024 | UL, MEC, APs (SA), UEs (SA) | Multiple | ✗ | ✓ | ✗ | ✓ | ✗ | ✗ | Known | Offloading volume, optimal paths and phase shifts of multiple RISs | SCA, SDR and AO |
| [117] | 2024 | UL, NOMA, MEC, SWIPT, UEs (SA), BS (MA), Relay (SA) | Single | ✗ | ✓ | ✗ | ✓ | ✗ | ✗ | Perfect | The RIS phases, the relay's power allocation factors, the offloading transmit power at the users, the offloading task ratio, and the PS ratio of the SWIPT | SCA & a penalty method |
| [118] | 2024 | UL, MEC, NOMA, BSs (MA), UEs (SA) | Multiple | ✗ | ✓ | ✓ | ✓ | ✗ | ✗ | Known | Equalizer, transmit power and RIS phase shift matrix | A new NOMA scheme based on IGIC |

[119] put forward a joint beamforming algorithm for RIS in multiuser MISO communications, using statistical CSI to reduce channel estimation overhead. The optimization employed the PPO algorithm, an actor-critic reinforcement learning method. Simulations showed that this approach improved user sum rates and outperformed existing methods by effectively integrating beamforming with statistical CSI. In another paper [120], the authors addressed optimal resource allocation for RIS-assisted dynamic WNs with uncertain and time-varying channels. They proposed a novel online RL-based approach, modeling the network as a state-space model. The resource allocation problem was formulated as a finite-horizon joint optimal control of users' transmit powers and RIS phase shifts. Using an Actor-Critic design with neural networks, the approach learned optimal resource allocation policies in real time. Numerical simulations confirmed the effectiveness of the scheme.

Similarly, the authors in [121] introduced a UAV-enabled multicast network with an added RIS to enhance service quality. They devised the Multi-Pass Deep Q Network (BT-MP-DQN) algorithm to jointly optimize UAV movement, RIS reflection matrix, and beamforming design to maximize sum rates. Simulation results validate its effectiveness and the benefits of deploying RIS, surpassing conventional multicast channels. In another work [122], the authors explored a RIS-assisted symbiotic radio system for short-packet transmissions, assessing ON/OFF reflection and binary phase shift keying at the RIS. It formulated an EE optimization problem, jointly optimizing transmit beamforming and RIS phase shifts. A DRL framework was developed to solve this, showing improved EE and convergence compared to existing methods in simulations. This paper [123] presented a low-complexity approach to optimizing phase shifts and transmit beamforming in a downlink MISO system using deep unfolding. Traditional iterative methods like BCD, which suffered from high latency and computational complexity, were replaced by an interpretable Neural Networks (NN) with few trainable parameters. This NN was trained offline and deployed online, reducing computational demands while maintaining comparable Weighted Sum Rate (WSR) performance, as confirmed by numerical results.

The authors in [124] investigated RIS-aided FD-MIMO communication for high spectral efficiency and EE. They optimized transmit and receive covariances and RIS phase-shift matrices for resource efficiency maximization. Two algorithms, DRL and AO, were proposed to solve the problem under varying channels. Simulations showed RIS-aided FD-MIMO's advantages, with the DRL algorithm achieving high resource efficiency with reduced complexity. This paper [125] introduced a framework integrating holographic MIMO and RIS for power-efficient beamforming in 6G wireless systems. They optimized system utility considering channel capacity, beampattern gains, distances, and losses. Using a Long Short-Term Memory (LSTM)-based for location determination and RL for resource allocation, they achieved significant power savings compared to baseline methods, enhancing holographic beamforming for user service. Another paper [126] explored resource allocation optimization in a multi-user WNs with RIS. It introduced a dynamic technique considering RIS hardware constraints and channel uncertainties. The online RL algorithm, combined with neural networks, learned optimal policies for power and phase shift control. Simulation results validated its effectiveness. Moreover, this paper [127] investigated spectrum sharing between an RIS-assisted cellular system and a MIMO radar. An optimization problem was formulated to balance communication and radar operations while minimizing interference. A meta-reinforcement learning (MRL) algorithm was developed to efficiently solve the problem, showing superiority in simulations over traditional methods by effectively controlling interference.

This paper [128] delved into dynamic optimization for resource allocation and RIS control in multi-cell Orthogonal Frequency-Division Multiple Access (OFDMA) systems. It proposed optimized subcarrier and power allocation, alongside RIS reflection element control. To achieve this, a Multi-agent DRL (MADRL) approach was introduced, utilizing DDQN and DDPG algorithms. Additionally, a transfer learning framework enhanced neural network adaptability, improving convergence speed and sum rate compared to baseline algo-rithms. Similarly, the authors in [129] proposed an AI and RIS-empowered communication system for Vehicular Ad-hoc Networks (VANETs) to tackle increasing traffic, delay sensitivity, and energy constraints. They designed a hybrid framework combining RIS-aided short-range and cellular communication for VANETs, including a head selection method for enhanced data transmission. Additionally, they developed a DRL scheme for efficient network resource control and allocation, focusing on EE and low latency. Their experiments validated the effectiveness of this system in improving VANETs's performance.

Unlike the uplink scenario in [128], [129], the authors in [130] focused on the downlink scenario and enhanced physical layer security in spectrum-sharing OFDM systems. DRL techniques, including Dueling Double Deep Q Networks (D3QN) and Soft Actor–Critic (SAC) algorithms, were applied to address non-convex optimization problems, resulting in improved security transmission rates compared to existing schemes. Different from single RIS-based frameworks in [128]–[130], The authors in [131] considered multiple active RISs and proposed a two-stage online RL algorithm for optimizing resource allocation in RIS-assisted Wireless Mobile Ad-hoc Networks (MANETs). Simulations validated its efficacy in improving network quality and optimizing resource allocation.

The authors in [132] explored over-the-air FL in a heterogeneous network and proposed the RIS-assisted Over-the-air Adaptive Resource Allocation for Federated learning (ROAR-Fed) algorithm. It optimized communication, computation, and learning resources with RIS assistance. Shown to be convergent, the algorithm effectively handled heterogeneous data and imperfect CSI. Overall, the study emphasized the benefits of RIS-assisted learning in this context. Similarly, in [133], the authors suggested combining RIS with mobile D2D communications to boost communication quality. They formulated and solved an optimization problem using a MADRL-based algorithm. Simulation results confirmed the approach's effectiveness in enhancing system performance, especially with more reflecting elements deployed. This paper [134] explored a UAV-RIS-assisted maritime communication system designed to combat jamming and optimize system EE. It introduced an adaptive energy harvesting scheme for simultaneous information transmission and energy harvesting. Using DRL, the study optimized BS power, UAV-RIS placement, and RIS beamforming. Simulations indicated that this approach outperformed existing methods in EE and harvesting under real-world conditions. This paper [135] explored a hybrid active-passive RIS-enabled multi-user communication system, optimizing RIS element scheduling, beamforming coefficients, and power allocation to maximize EE. Using Dinkelbach relaxation, the complex optimization problem was simplified and solved with AO. An exhaustive search set the optimal RIS mode, while a Big-M formulation reduced complexity. Numerical results demonstrated that the hybrid RIS system surpassed traditional systems in EE, particularly when elements were active.

Distinct from all the aforementioned OMA-based works [128]–[131], this paper [136] considered NOMA and examined enhancing FL in IoT networks with RIS to improve model aggregation efficiency. It aimed to minimize train-ing latency while considering IoT device energy constraints. Investigating Frequency-Division Multiple Access (FDMA) and NOMA protocols, it optimized RIS phase shifts, communication scheduling, and IoT device parameters. Efficient algorithms were developed, showing substantial latency reduction with RIS-assisted FL systems, especially with NOMA-based aggregation, outperforming FDMA in training latency. Another paper [137] explored a D2D system with NOMA and multiple RISs to enhance service quality and NOMA gains. RISs notably increased the average sum data rate for D2D groups. The study used a MAMP-DQN framework to optimize sub-channel assignments, power allocation, and phase shifts. Results showed a 27% higher rate for NOMA-empowered groups compared to OMA, with more enhancements by increasing RISs or reflective elements.

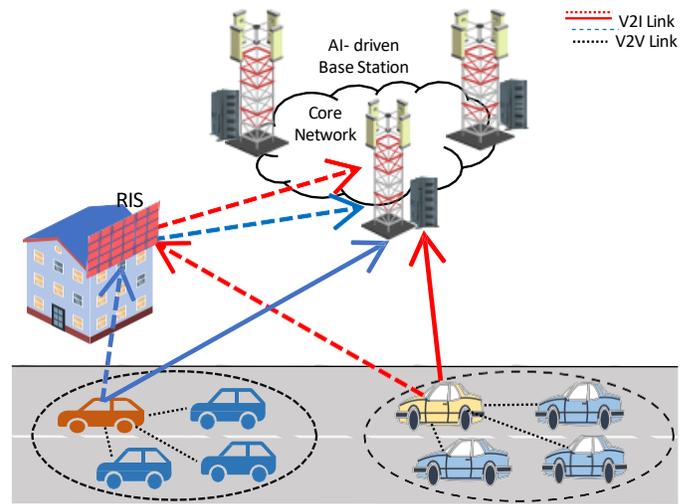

Fig. 11: AI and RIS empowered vehicular communication network

The authors in [138] introduced a DRL-based method for optimizing hybrid beamforming, phase shifts, and power allocation in RIS-assisted mmWave communication. Experiments showed a minimum 37% higher sum rate compared to baseline algorithms. The authors in [139] focused on the downlink scenario and proposed integrating IoT applications into Beyond Fifth Generation (B5G) networks with a RIS-assisted wideband THz communication system. They optimized RIS element reflection coefficients, BS transmit power, and wideband THz resource block allocation. Using a supervised learning strategy, simulation results showed significant spectral efficiency gains for enhanced mobile broadband (eMBB) services, up to 49% with an 11x11 RIS. Additionally, the ensemble learning model proved efficient for real-time resource management. Recent advancements [140] focused on using THz communications and RISs for enhancing wireless VR applications. It developed a Semi-Markov Decision Process model to optimize THz path allocation, considering the impact of RIS hardware failures. An iterative algorithm was proposed to ensure reliable connections and maximize system rewards, providing a basis for stable VR services in THz RIS networks despite potential hardware issues. VR applications in downlink THz MIMO small BS systems utilize statistical CSI for RIS path allocation via a Semi-Markov decision process. In scenarios with known CSI, downlink THz MIMO configurations optimize precoding, phase shifts, and adaptive sub-band bandwidth using MHGphormer learning algorithms [141].

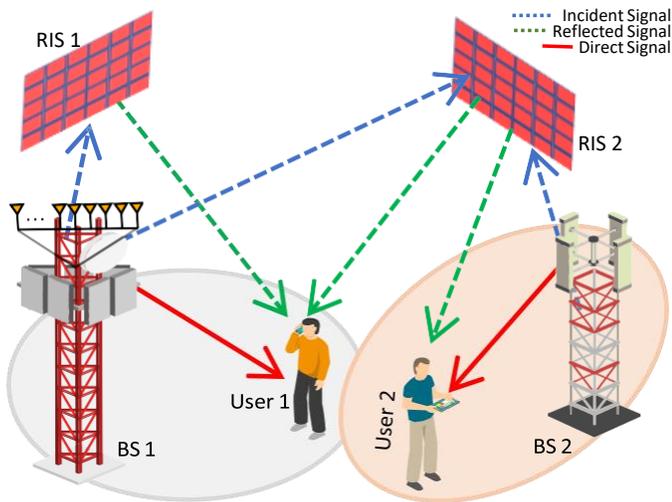

Fig. 12: RIS Based Multi-Cell DL MISO System

Another article [142] explored a STAR-RIS-assisted Vehicle-to-Everything (V2X) communication system to enhance coverage in urban environments. The study optimized data rates for V2I users and ensured latency and reliability for V2V communications by adjusting spectrum allocation, STAR-RIS element settings, and power levels. It employed a Markov Decision Process and a Double Deep Q-Network (DDQN) with an attention mechanism to solve complex optimization challenges, demonstrating significant improvements over conventional methods in numerical tests.

[143] examined optimal placement and resource allocation in a multi-UAV RIS-assisted WNs. They proposed an online RL-based algorithm integrating Deep Q-Learning (DQL) for UAV deployment and actor-critic RL for resource allocation. Simulations its effectiveness. In [144], the authors explored STAR-RIS with perfect CSI and integrating UAVs for MEC to enhance IoT communications, focusing on minimizing energy consumption. Tackled the optimization of task offloading, STAR-RIS trajectory, and power settings using Proximal Policy Optimization (PPO). Simulations showed this method effectively reduced energy usage compared to existing solutions. Another paper [145] proposed a learning-based approach to optimize UAV trajectory and RIS reflection coefficients in UAV-RIS-assisted cognitive Non-Terrestrial Networks (NTNs), aiming to enhance secrecy performance. The method accounted for practical RIS phase shifts, outdated CSI, and satellite interference. A DRL algorithm optimized the UAV trajectory, while a Double Cascade Correlation Network (DCCN) adjusted the RIS coefficients. Simulations demonstrated significant improvements in secrecy performance.

[146] proposed a Federated Spectrum Learning (FSL) framework, merging RISs with FL for spectrum sensing. RISs adopted spectrum learning via Convolutional Neural Network (CNN) models, tackling challenges in phase shifts configuration, user-RIS association, and bandwidth allocation.

They system utility while considering FL prediction accuracy. Simulation results demonstrated FSL's benefits in spectrum prediction accuracy and system utility, particularly with increased RISs and reflecting elements.

In [147], the authors proposed an RIS-assisted edge heterogeneous network for MEC to minimize latency and energy use. Their network included macro and small BSs with MEC servers, leveraging RIS for additional computation offloading. They aimed to reduce long-term energy consumption while meeting QoS and resource limits. Using a two-timescale mechanism combining matching theory and DRL, they addressed the optimization challenge. Simulations verified its effectiveness in enhancing network performance. In another paper [148], the authors addressed improving MEC for IoT devices using multiple RISs across different areas. They proposed a cooperative multi-agent RL model for resource management, achieving reduced latency and efficient resource utilization. Simulations demonstrated up to 11.84% reduction in computation resource needs compared to existing methods, highlighting scalability with increasing IoT device demands.

Unlike OMA-based MEC works [146]–[148], the authors in [149] considered NOMA and investigated the integration of covert communications with MEC in 6G networks. They RIS and NOMA technologies to ensure secure transmission and conceal the presence of stronger signal receivers. By analyzing the system's performance and proposing an RL-based power-allocation optimization algorithm, the authors demonstrated the effectiveness of their approach in achieving improved connectivity and covert communication. The authors in [150] introduced an algorithm for computation offloading in RIS-enhanced MEC networks. It optimizes RIS parameters, offloading decisions, and resource allocation to reduce delay, energy consumption, and operator costs. Comparative evaluations confirm its superiority over non-RIS learning algorithms and classical approaches. This paper [151] integrated RIS into a MEC framework, proposing a joint computation offloading and power allocation strategy using NOMA with RIS assistance. Employing an Markov decision process (MDP) model and DDQN algorithm, they aimed to optimize computation offloading and minimize system costs. Simulation results confirmed reduced offloading delay and lower costs, with satisfactory convergence. Similarly, the paper [152] introduced a joint task offloading and resource allocation strategy for UAV-MEC networks with RIS. Using MDP modeling and DDQN, they optimized to minimize UAV energy consumption and enhance QoS. Simulation results validate its effectiveness over existing benchmarks. Similarly, the authors in [153] developed a novel algorithm for efficient edge inference in 6G networks with RISs. It combined Lyapunov stochastic optimization with DRL to optimize data compression, resource allocation, and RIS reflectivity. The dynamic learning algorithm prioritized energy-efficient edge classification while adapting to varying conditions. Simulations validated its effectiveness in balancing energy, delay, and accuracy. In another paper [154], the authors presented a system that combined coordinated multipoint (CoMP) and RISs to optimize 3D video rendering in wireless VR networks using collaborative MEC servers. The authors developed a hybrid learning framework integrating

DRL with AO for beamforming. The proposed method effectively reduced power consumption and enhanced transmission efficiency.

This article [155] explored enhancing task offloading security in MEC-enabled Industrial IoT networks by deploying RIS to shield against eavesdroppers. The study developed an optimization framework to maximize secrecy computation efficiency, involving RIS phase shifts, power control, and time-slot allocation. A DRL algorithm using Deep Deterministic Policy Gradient (DDPG) was employed to solve this complex problem, with numerical results showing improved security performance compared to baseline methods.

*Summary and Lessons Learned:* **Summary:** Table IX provides a comprehensive overview of AI-driven resource allocation frameworks in RIS-aided systems, highlighting diverse network types such as SISO, MISO, MIMO, HetNets, NOMA, mmWave, THz, VC, UAV, and edge networks. These studies predominantly focus on optimizing critical design variables, including power allocation, phase shifts, beamforming, and UAV trajectories, with most frameworks assuming known CSI and a few addressing perfect or statistical CSI scenarios.

Single RIS setups are commonly explored to enhance system performance, while multiple RIS configurations are addressed in studies such as [124], [131], [137], [138], [140], [143], [146], [148], [150]. These works optimize key design variables like power allocation, phase shifts, and beamforming by leveraging both known and statistical CSI, demonstrating significant improvements in network efficiency and user experience. Active RIS configurations, discussed in studies such as [127]–[131], [134], [146], [148], [150], focus on optimizing transmission power, computing resources, and RIS phase shifts, further highlighting the versatility and potential of RIS in complex network scenarios.

Despite these advancements, the complexity of managing multiple RIS configurations presents significant computational challenges, especially when optimizing interconnected design variables like power allocation, phase shifts, beamforming, and channel assignments. A critical gap exists in understanding the impact of imperfect CSI on these optimization strategies. Addressing this gap is crucial for enhancing the robustness and adaptability of RIS frameworks in real-world deployments. Aerial RIS configurations introduce additional complexity due to the dynamic nature of airborne platforms, requiring precise control over RIS settings in fluctuating environments. Optimizing trajectory, phase shifts, and power allocation in such scenarios remains a significant challenge, particularly under varying conditions and imperfect CSI. The frequent use of advanced AI techniques, such as DRL and its variants (Actor-Critic, DDQN, PPO), underscores their effectiveness in tackling complex decision-making processes in dynamic settings. However, further research is needed to refine these methods and develop more robust, adaptive resource allocation strategies, especially for large-scale, dynamic, and multi-RIS environments.

**Lessons Learned:**

· Enhancing System Performance and Service Delivery: Autonomous UAV-assisted networks increasingly depend on AI to improve system performance and service delivery [121], [122]. AI, particularly through deep learning, is essential in areas such as emergency response, event coverage, and providing services in remote regions. These AI-driven methods enhance real-time data processing, optimize resource allocation, and improve flight path planning. As a result, integrating AI into UAV networks leads to more efficient and effective service delivery, allowing these systems to quickly adapt to changing conditions and provide accurate responses, even in challenging environments.

· Task Allocation and System Efficiency: UAVs face significant challenges in optimizing task allocation, computation, and real-time decision-making. However, leveraging AI can substantially improve their capabilities [122], [125]. AI-driven models enable predictive control, trajectory optimization, resource allocation, and conflict resolution, which collectively enhance system efficiency. These improvements lead to reduced energy consumption, optimized resource utilization, and increased safety, allowing UAVs to perform more effectively in complex and dynamic environments.

· Embracing Innovation for Enhanced System Performance: Our literature review reveals that traditional methods are increasingly inadequate, with AI-integrated algorithms demonstrating superior performance [124], [154]. It is imperative for researchers to choose algorithms that are specifically tailored to their network scenarios and to move away from outdated techniques that may lead to inefficiencies. By embracing innovative and context-specific solutions, researchers can significantly enhance system performance and efficiency, thereby reducing the time and resources typically consumed by conventional methods.

· Enhancing Security Against Eavesdropping: While MEC offers many advantages, wireless task offloading introduces security risks due to the broadcast nature of electromagnetic signals, making IoT devices susceptible to eavesdropping. Integrating RIS with PLS techniques can bolster security by enhancing signal strength at intended receivers and reducing the likelihood of signal leakage to potential eavesdroppers [149], [155].

## IV. OPEN ISSUES AND FUTURE RESEARCH DIRECTIONS

As we progress towards 6G networks, the potential of RIS to enhance wireless communication is increasingly recognized. RIS technology not only boosts signal strength but also supports complex functions such as channel estimation and user localization. While current research has addressed many issues, there remain open questions and challenges that researchers need to tackle. This section will explore the open issues and challenges facing RIS in various environments such as THz communications, vehicular networks, and UAV communications, outlining the future directions for this promising technology in the evolution of WNs.

The discussion delves into the following open issues and significant challenges:

· **Strategic RIS Deployment:** The deployment of RIS is essential for enhancing wireless network performance,

TABLE IX: Summary of AI-based Resource Allocation Frameworks

| Ref | Year | System Model | RIS Details | | | | | | | CSI | Design and Optimization | |
|---|---|---|---|---|---|---|---|---|---|---|---|---|
| | | | No's | A | P | H | R | T | STAR | | Designed Variable | Optimization Methodology |
| [119] | 2024 | DL, SIMO, BS (MA), UEs (SA) | Single | ✗ | ✓ | ✗ | ✓ | ✗ | ✗ | Statistical | Beamforming vectors at the BS and phase shifts at the RIS | The PPO algorithm, a well-established actor-critic-based DRL approach |
| [120] | 2022 | DL, MISO, BS (MA), UEs (SA) | Single | ✗ | ✓ | ✗ | ✓ | ✗ | ✗ | Not required | UEs transmit power & RIS phase shift. | RL |
| [121] | 2023 | DL, MISO, UAV (MA), UEs (SA) | Single | ✗ | ✓ | ✗ | ✓ | ✗ | ✗ | Known | The UAV movement, RIS reflection matrix, and beamforming design from the UAV to users | DQN |
| [122] | 2024 | DL, MISO, BS (MA), UEs (SA) | Single | ✗ | ✓ | ✗ | ✓ | ✗ | ✗ | Known | Transmit power, beamforming and phase shift | DRL |
| [123] | 2024 | DL, MISO, BS (MA), UEs (SA) | Single | ✗ | ✓ | ✗ | ✓ | ✗ | ✗ | Known | Transmit beamforming and phase shift | NN |
| [124] | 2023 | DL, MIMO, BS (MA), UEs (MA) | Single | ✗ | ✓ | ✗ | ✓ | ✗ | ✗ | Known | Transmit covariance, optimal receive covariance, and phase-shift | DRL & AO |
| [125] | 2023 | DL, MIMO, BS (MA), UEs (MA) | Single | ✗ | ✓ | ✗ | ✗ | ✗ | ✓ | Sensory | Power | Actor-Critic based DDPG |
| [126] | 2023 | DL, MIMO, BSs (MA), Receivers (SA) | Single | ✗ | ✓ | ✗ | ✓ | ✗ | ✗ | Known | Power and RIS phase shift | A novel online data-enabled actor-critic-barrier RL algorithm |
| [127] | 2024 | DL, MIMO, BS (MA), UEs (MA) | Single | ✓ | ✗ | ✗ | ✓ | ✗ | ✗ | Known | The communication transmit precoder matrix, RIS phase shift matrix, and transmit waveform of radar | A low-complexity meta-reinforcement learning (MRL) algorithm based on DRL |
| [128] | 2022 | UL, OFDMA, BS (MA), UEs (SA) | Single | ✗ | ✓ | ✗ | ✓ | ✗ | ✗ | Known | Amplitude and phase shift of all reflection elements | MADDQN & MADDPG |
| [129] | 2022 | UL, VANET, V2I, V2V, Edge server, BSs (MA), Vehicles (MA) | Single | ✗ | ✓ | ✗ | ✓ | ✗ | ✗ | Known | Power, RIS-reflection phase shift, and BS detection matrix | DRL |
| [130] | 2023 | DL, OFDM, BSs (MA), UEs (SA), Eves (SA) | Single | ✗ | ✓ | ✗ | ✓ | ✗ | ✗ | Known | The beamforming of SBS, the RIS's reflecting coefficient and the channel allocation | DRL, D3QN and SAC |
| [131] | 2023 | MANET, Transmitters (MA), Receiver (MA) | Multiple | ✓ | ✗ | ✗ | ✓ | ✗ | ✗ | Known | Power and phase shift | An online distributed cooperative actor-critic RL & NN |
| [132] | 2023 | UL, HetNets, BS (SA), Edge devices (SA) | Single | ✗ | ✓ | ✗ | ✓ | ✗ | ✗ | Known | Power and RIS phase shift | ROAR-Fed based on FL |
| [133] | 2023 | UL and DL, HetNets, BS (MA), UEs (SA) | Single | ✗ | ✓ | ✗ | ✓ | ✗ | ✗ | Perfect | Mode selection, channel assignment, power allocation, and discrete phase shift selection | MADRL-based framework combines both the multi-pass deep Q-networks (MP-DQN) algorithm and the decaying DQN algorithm |
| [134] | 2024 | DL, HetNets, BS (MA), UEs (SA), Jammer | Single | ✗ | ✓ | ✗ | ✓ | ✗ | ✗ | Known | The BS transmit power, placement of UAV-RIS, and RISs reflecting beamforming | An intelligent resource management approach based on DRL |
| [135] | 2024 | DL, HetNets, BS (MA), UEs (SA) | Single | ✗ | ✓ | ✗ | ✓ | ✗ | ✗ | Perfect | The RIS element scheduling and beamforming coefficients, power allocation coefficients | Dinkelbach and AO |
| [136] | 2022 | UL, NOMA, BS (SA), IoT devices (SA) | Single | ✗ | ✓ | ✗ | ✓ | ✗ | ✗ | Perfect | RIS phase shifts, communication resource, transmit power, frequencies | FL |
| [137] | 2023 | UL, NOMA, BS (MA), UEs (MA), D2D groups | Multiple | ✗ | ✓ | ✗ | ✓ | ✗ | ✗ | Known | Phase shift, sub-channel assignment and power allocation | MAMP-DQN & the formulated Markov game (MG) |
| [138] | 2023 | DL, mmWave, BS (MA), UEs (SA) | Single | ✗ | ✓ | ✗ | ✓ | ✗ | ✗ | Known | Hybrid beamforming, phase shifts, and power allocation | DRL |
| [139] | 2023 | DL, THz, BS (MA), eMBB & URLLC UEs (MA) | Single | ✗ | ✓ | ✗ | ✓ | ✗ | ✗ | Known | Reflection coefficient and the transmit power | DL & ensemble learning |
| [140] | 2023 | DL, THz, MIMO, Small BS (MA), VR UEs (SA) | Multiple | ✗ | ✓ | ✗ | ✓ | ✗ | ✗ | Known | RIS path allocation | A Semi-Markov decision Process (SMDP)-based path allocation, and an optimal iterative algorithm |
| [141] | 2024 | DL, THz, MIMO, BS (MA), UEs (SA) | Single | ✗ | ✓ | ✗ | ✓ | ✗ | ✗ | Known | The precoding, phase shifts, and Adaptive sub-band bandwidth | An unsupervised MHGphormer learning algorithm based on NN |
| [142] | 2024 | UL, V2X, MEC, BS (MA), VUEs (SA) | Single | ✗ | ✓ | ✗ | ✗ | ✗ | ✓ | Perfect | The spectrum allocation, amplitude and phase shift values of STAR-RIS | Markov Decision Process, DDQN, a standard optimization-based approach |
| [143] | 2023 | DL, UAVs, BS (MA), UEs (SA) | Multiple | ✗ | ✓ | ✗ | ✓ | ✗ | ✗ | Known | Power and RIS phase shift | A DQL based K-means clustering algorithm & an online actor-critic RL algorithm |
| [144] | 2024 | UL, UAV, MEC, BS (MA), IoT devices | Single | ✗ | ✓ | ✗ | ✓ | ✗ | ✗ | Perfect | Task offloading, trajectory, amplitude and phase shift coefficients, and transmit power | DRL and PPO |
| [145] | 2023 | DL, UAV, Satellites, BS (MA), Eve (MA), signal receivers | Single | ✗ | ✓ | ✗ | ✓ | ✗ | ✗ | Outdated | The UAV trajectory and RIS reflection coefficients | DRL and double cascade correlation network (DCCN) |
| [146] | 2022 | UL, Edge server, BS (MA), UEs (MA) | Multiple | ✗ | ✓ | ✗ | ✓ | ✗ | ✗ | Perfect | RISs phase shifts, UE-RIS association, and wireless bandwidth | FSL |
| [147] | 2022 | UL, MEC, BS (SA), UEs (SA) | Single | ✗ | ✓ | ✗ | ✓ | ✗ | ✗ | Known | RIS phase shift | Matching theory and DRL |
| [148] | 2022 | DL, MEC, BS (MA), IoT devices (SA) | Multiple | ✗ | ✓ | ✗ | ✓ | ✗ | ✗ | Known | Communication and computation resources | A multi-agent actor-critic method with an attention mechanism |
| [149] | 2023 | DL, MEC, NOMA, Tx (SA), Rxs (SA), Jammer (SA) | Single | ✗ | ✓ | ✗ | ✓ | ✗ | ✗ | Known | Power | RL |
| [150] | 2023 | UL, MEC, Small BSs (MA), MD | Multiple | ✓ | ✗ | ✗ | ✓ | ✗ | ✗ | Known | The phase shift and amplitude of RIS, offloading decision, and MEC resource allocation strategy | DDPG |
| [151] | 2023 | UL, MEC, NOMA, AP (MA), UEs (MA) | Single | ✗ | ✓ | ✗ | ✓ | ✗ | ✗ | Known | Power and RIS phase shift | Lyapunov, MDP and DDQN |
| [152] | 2023 | UL, MEC, UAV, AP (MA), UEs (SA) | Single | ✗ | ✓ | ✗ | ✓ | ✗ | ✗ | Known | Task offloading decisions, allocation of UAVs' computing resources, communication resource allocation, and phase shift of RIS | MDP & DDQN |
| [153] | 2023 | UL, MEC, AP (MA), MDs (MA) | Single | ✗ | ✓ | ✗ | ✓ | ✗ | ✗ | Known | Data compression, resource allocation, and RIS reflectivity | Lyapunov stochastic optimization & DRL |
| [154] | 2024 | DL, MEC, BSs (MA), UEs (SA) | Single | ✗ | ✓ | ✗ | ✓ | ✗ | ✗ | Known | The video caching and rendering and the beamforming for both BSs and RIS | DRL, DDQN-AO |
| [155] | 2024 | UL, MEC, AP (MA), IoT devices (SA), Eves (SA) | Single | ✗ | ✓ | ✗ | ✓ | ✗ | ✗ | Known | RIS phase shift, power control, local computation rate, and time-slot allocation | DDPG in DRL |

especially by improving signal coverage and spectral efficiency in a cost-effective manner. RIS components can intelligently manipulate signal propagation, significantly enhancing network reliability and data throughput, particularly in scenarios where traditional infrastructure is inadequate. Researchers should tailor strategic deployment plans for RIS based on the unique needs of different environments to maximize coverage and connectivity [39], [137], [156]. The benefits of deploying RIS include improved signal quality and coverage, increased network efficiency, reduced infrastructure costs, and scalable, flexible network adaptability. To effectively deploy RIS, strategies should focus on developing sophisticated channel estimation methods, AI-driven dynamic control systems, comprehensive deployment strategies that consider geographical and architectural factors, and ensuring compatibility with existing network infrastructure to facilitate seamless integration and operation, thereby fostering the development of robust, efficient, and cost-effective wireless networks [30], [157].

- **Advanced Channel Estimation and Mobility Adaptation:** Effective channel estimation is crucial for optimizing RIS performance, particularly challenging in high-mobility environments such as vehicular and UAV networks where the dynamic and unpredictable nature complicates real-time CSI tracking, essential for maintaining robust communication links [158], [159]. The passive nature of RIS and rapid movements exacerbate these challenges, often rendering current methods inadequate. To address these issues, future research should focus on AI-driven approaches, such as machine learning-based channel prediction, which leverages historical data to forecast future channel states, allowing for proactive adjustments. Additionally, dynamic phase adaptation, which involves real-time adjustments of the phase settings on RIS units to instantly react to changes in the channel, and novel sensing techniques that use advanced sensors to measure environmental factors affecting signal propagation, can significantly enhance the accuracy and efficiency of channel estimation. Moreover, adapting to high mobility necessitates innovative strategies that ensure reliable connections and optimize network performance in fast-moving scenarios, including the development of mobility-aware algorithms that account for the predicted trajectories of mobile users and the use of cooperative RIS networks, where multiple RIS units communicate and coordinate to seamlessly hand over the signal reflection duties as mobile users move through different network zones [160].

· **Energy Efficiency Optimization and Resource Management:** EE remains a critical challenge in RIS-assisted networks, particularly as the demand for high-speed, high-capacity communication intensifies. The burgeoning need for robust wireless communication systems across various configurations, such as SISO, MIMO, NOMA, and edge networks, underscores the urgency to develop more sophisticated methods to ensure sustainable network operations. This is crucial, as the growing complexity of these networks increases the potential for significant energy consumption, especially when leveraging the capabilities of RIS to enhance communication quality. To effectively address these challenges, it is essential to focus on the development of intelligent algorithms for both beamforming and power management. These algorithms must dynamically adjust to fluctuating network conditions and user demands to optimize energy use while maintaining high communication standards. Specifically, beamforming algorithms should be designed to adjust the phase and amplitude of signals with precision, minimizing energy waste by focusing energy delivery to necessary areas without superfluous spread. Additionally, in MEC environments, the challenge of optimizing energy consumption is exacerbated by the dual demands of processing power for task offloading and the energy requirements of RIS operations. This calls for innovative energy management strategies that can intelligently balance the trade-off between the energy saved through efficient task offloading to edge servers and the energy expended in adjusting RIS elements to support this offloading [113], [150]. Such strategies should include refining algorithms that manage energy use more effectively and developing new technologies or methodologies that reduce the inherent energy consumption of RIS operations. Moreover, future research should explore the integration of AI and ML to predict network conditions and user behaviors more accurately. This approach would enable networks to pre-emptively adjust settings to minimize unnecessary energy consumption. Additionally, the incorporation of renewable energy sources directly into the RIS infrastructure could provide a sustainable way to power these devices, thus enhancing the overall EE of the network. Exploring these advanced strategies is essential for evolving toward more sustainable RIS-assisted networks that can meet the growing demands of modern telecommunications without compromising environmental integrity.

· **Addressing Security Risks in RIS-based Networks:** As RIS technology becomes increasingly embedded in mobile and edge networks, particularly in contexts like vehicular and UAV systems, it introduces a new spectrum of security risks. These risks stem from the inherent capabilities of RIS to alter wireless communication environments, making them susceptible to unauthorized interception, signal manipulation, and other forms of malicious attacks. The primary challenge lies in the passive nature of RIS, which, while reducing power consumption and complexity, significantly limits its ability to autonomously detect and counter security threats. This vulnerability necessitates the development of robust security frameworks tailored to the unique operational paradigms of RIS-assisted networks. Effective protection strategies could include the implementation of advanced encryption methods that safeguard data as it passes through RIS-modified paths, the establishment of secure protocols for configuring RIS elements to prevent unauthorized access, and the deployment of real-time monitoring systems capable of dynamically detecting and responding to potential security breaches. Moreover, integrating AI and ML can significantly bolster these security measures by enabling networks to automatically detect anomalies and adapt to evolving threats, thereby maintaining the integrity and security of communication channels. This comprehensive approach is critical not only for mitigating the security vulnerabilities introduced by RIS technology but also for ensuring that its deployment enhances network capabilities without compromising security [161], [162].

· **Enhancing Environmental Resilience in Mobile and UAV Applications:** The integration of RIS into mobile and UAV networks is significantly challenged by environmental factors such as wind, vibrations, and physical obstacles, which can disrupt the alignment of RIS elements and lead to degraded signal quality and unreliable communication. For UAVs, this issue is particularly severe as their constant movement, combined with environmental variables, can result in frequent misalignment and unstable connections. To effectively mitigate these problems, it is essential to develop RIS designs that are robust enough to withstand environmental stressors while maintaining optimal performance. Key to this resilience is the implementation of adaptive beamforming technology, which can adjust in real-time to changing conditions, thereby ensuring stable and reliable communication. Additionally, the integration of sensors within RIS units could significantly enhance system responsiveness by detecting and adapting to environmental changes in real-time. Further research should focus on developing predictive models that can anticipate environmental impacts on RIS configurations, allowing for preemptive adjustments that maintain communication quality despite challenging conditions [33].

· **Hardware Design and Implementation Complexities:**

The design and implementation of RIS hardware present fundamental challenges, particularly in advanced communication environments such as millimeter-wave and terahertz systems. Optimizing the physical dimensions and configurations of RIS for specific applications, whether they are intended for indoor SISO/MIMO systems, outdoor vehicular networks, or large-scale infrastructural deployments, demands meticulous consideration of performance, cost, and EE [157], [163]. Future hardware designs need to prioritize creating compact, modular, and reconfigurable RIS units that can seamlessly adapt to various scenarios without introducing excessive complexity. This approach necessitates a balance between physical robustness and flexibility to accommodate the unique demands of different environments and use cases. Furthermore, integrating AI-driven control mechanisms can significantly enhance the functionality and adaptability of RIS hardware, enabling more intelligent and responsive operations within the network. Such advancements would not only streamline the deployment of RIS technologies but also bolster their efficiency and effectiveness in dynamic communication landscapes [31], [33], [39], [157], [159], [162]–[164].

· **Sensing and Communication:** The growth of IoT, IoV, and UAV networks underscores the critical need for integrating communication and sensing in future WNs, a key focus of the IMT-2020 group for 5G-Advanced and 6G networks. RIS is pivotal in this integration, offering promising solutions to enhance both communication efficiency and sensing accuracy. However, several challenges remain in effectively applying RIS for both communication and positioning. These include the need for sophisticated resource allocation strategies, managing signal interference, developing suitable network architectures, and optimizing RIS coefficients to tailor the propagation environment effectively. Addressing these challenges requires a holistic approach that considers the complex interplay between communication and sensing functionalities, ensuring that advancements in RIS technology contribute meaningfully to the robustness and efficiency of future wireless networks [158].

· **Near-field Localization:** Effectiveness is significantly impacted by path-loss in its reflected phases, necessitating either proximity to transmitters or larger sizes to achieve higher Signal-to-Noise Ratios (SNR). However, as RIS sizes and operating frequencies increase, the near-field range also extends, as indicated by the Fraunhofer distance [165]. This extension challenges the accuracy of the far-field, plane wave assumptions prevalent in many models and suggests a pressing need for the adoption of models that incorporate both spherical and plane wave components. This need is particularly acute in applications such as indoor localization and IoT positioning, where the increased RIS sizes and frequencies make traditional modeling approaches less effective. To address these challenges, future research should focus on developing and refining hybrid wave models that can more accurately represent the complex waveforms encountered in these scenarios, ensuring that RIS technologies continue to provide reliable and precise positioning capabilities as they scale in size and function [158].

· **Holographic Surfaces:** Holographic surfaces are becoming increasingly prominent in RIS-assisted wireless communications, characterized by a densely integrated array of tiny reflecting elements that create a more continuous aperture. This innovative design significantly enhances signal directionality and reduces sidelobe leakage compared to traditional RIS configurations, offering improved performance in directing and shaping electromagnetic waves. Research, such as in [164], has begun to explore channel models, beamforming schemes, and system optimization specifically for holographic RIS, particularly within MIMO systems. However, key challenges persist, including managing the complex CSI required for these large element arrays. Additionally, there is a critical balance that must be struck in spacing these elements to optimize performance without exacerbating mutual coupling and signal correlation, which can degrade the overall system efficiency. Addressing these challenges is crucial for leveraging the full potential of holographic surfaces in enhancing wireless network capabilities and achieving superior communication system performance.

· **Digital Twins:** Digital twins are essential for bridging the gap between digital systems and physical spaces by creating digital replicas of physical objects, machines, and devices on a server. These replicas utilize both real-time and historical data to accurately mirror their physical counterparts, facilitating enhanced communication and interaction between the physical and digital worlds. This technology not only optimizes the operation of physical systems but also plays a crucial role within the context of 6G WNs. In 6G, digital twins can significantly aid in the optimal configuration and automation of RIS, a pivotal technology for the upcoming generation of networks. This integration allows for more precise control and optimization at various levels of network operation. Furthermore, digital twins contribute to the development of practical RIS solutions, enhancing the security, privacy, and overall performance of RIS-assisted 6G systems. By enabling the RIS controller to be trained in a virtual environment against various attack scenarios, digital twins ensure that the communication remains secure and stable, addressing both current and emerging security challenges effectively [161].

· **Quantum Communication:** Quantum communication is a promising technology for enhancing the reliability and security of data transmission in 6G networks. It uniquely offers high-level security because any eavesdropping or replication attempts in quantum communication immediately alter the quantum state, alerting the recipient to potential interference. While quantum communication theoretically promises secure long-distance communication, it can't solve all privacy and security issues. Implementing quantum cryptography faces challenges like errors and fiber attenuation over long distances. Various advanced techniques and quantum encryption methods, including quantum dense coding, teleportation, secure direct communication, secret sharing, and key distribution, are being explored to strengthen quantum communication security. Recent studies have particularly focused on quantum key distribution models to safeguard key security. Integrating quantum technology into RIS-assisted 6G networks has the potential to significantly improve communication quality beyond the capabilities of conventional systems. However, this field is still in its early stages, requiring extensive research and development [161].

## V. Conclusion

Our survey has provided an in-depth exploration into the innovative world of resource allocation within RIS-aided networks, underscoring the significant role of RIS in transforming network operations. We have meticulously analyzed a variety of RIS configurations — active, passive, and hybrid — and their functional modes, effectively linking theoretical RIS concepts with their practical implications in enhancing network performance. Key to our study is the detailed investigation of RIS applications across a broad spectrum of network architectures. This includes established frameworks like SIMO, MISO, and MIMO, as well as more complex networks such as heterogeneous wireless systems, THz communication networks, VC channels, and UAV-based communications. Here, RIS technology emerges as a beneficial tool and a critical component in optimizing resource allocation, offering versatility and efficiency in various network settings. One of the most compelling aspects of our survey is exploring how RIS technology converges with other advanced technologies, especially AI. This intersection presents an exciting frontier, promising to advance network efficiency further through AI-enhanced RIS strategies. It highlights a path towards more intelligent and responsive network systems, where resource allocation is dynamically optimized to meet evolving demands.

In concluding this comprehensive survey, it becomes clear that RIS technologies represent a paradigm shift in network resource management. While our research has illuminated many facets of RIS in network systems, it also points to an array of future challenges and unexplored avenues. These range from the need for advanced system modeling and the integration of RIS in emerging network technologies to the refinement of AI-driven algorithms for optimal resource allocation. Thus, this survey not only serves as a rich resource for current and future researchers in this field but also acts as a catalyst for further innovation in the domain of RIS-aided network resource allocation. As the landscape of network demands continues to evolve, the role of RIS in meeting these new challenges becomes increasingly pivotal, pointing towards a future where network systems are more efficient, adaptable, and intelligent.

## Acknowledgements

This work was supported by the Multimedia University Research Fellow Grant (MMUI/240021) and the TM Research and Development Grant (RDTC/241149).

## Conflict of Interests

The authors declare no conflicts of interest.